\documentclass[epj]{svjour}

\usepackage{epsfig}
\usepackage{amsmath}
\usepackage[usenames]{color}
\usepackage{ulem} 

\newcommand{\be}{\begin{eqnarray}}
\newcommand{\ee}{\end{eqnarray}}
\newcommand{\non}{\nonumber \\}
\newcommand{\po}{{\rm P}}
\newcommand{\npo}{{\rm NP}}

\newcommand{\ezp}{E_{0+}}
\newcommand*{\ket}[1]{|#1\rangle}
\newcommand*{\bra}[1]{\langle#1|}
\newcommand*{\braket}[1]{\langle#1\rangle}
\newcommand*{\fs}[1]{{#1\!\!\!/}}

\begin{document}

\title{\thanks{FZJ-IKP-TH-2009-19}The phase and pole structure of the $N^*(1535)$ in $\pi N\to\pi N$ and $\gamma N\to \pi N$}

\author{M. D\"oring\inst{1,2}\and K. Nakayama\inst{2,1}}
\institute{
Institut f\"ur Kernphysik and J\"ulich Center for Hadron Physics,
Forschungszentrum J\"ulich, D-52425 J\"ulich, Germany \thanks{\email{m.doering@fz-juelich.de}} \and Department of Physics and Astronomy, University of Georgia,
Athens, GA 30602, U.S.A. \thanks{\email{nakayama@uga.edu}}}

\abstract{
The nature of some baryonic resonances is still an unresolved issue. The case of the $N^*(1535)$ is particularly interesting in this respect due to the nearby $\eta N$ threshold and interference with the $N^*(1650)$. The $N^*(1535)$ has been described as a threshold effect, as a genuine 3-quark resonance, or as dynamically generated from the interaction of the octet of baryons with the octet of mesons. In the scheme of dynamical generation, predictions for the interaction of the $N^*(1535)$ with the photon can be made. In this study, we simultaneously analyze the role of the $N^*(1535)$ in the $\pi N\to\pi N$ and $\gamma N\to \pi N$ reactions and compare to the respective amplitudes from partial wave analyses. This test is very sensitive to the meson-baryon components of the $N^*(1535)$.  
}

\maketitle

\section{Introduction}
\label{sec:intro2}
The unitary extensions of chiral perturbation theory\\ $(U\chi PT)$ have brought a new light to the meson-baryon interaction, showing that some well-known resonances qualify as being dynamically generated. In this picture, the Bethe-Salpeter resummation of elementary interactions, derived from chiral Lagrangians, guarantees unitarity and leads, at the same time, to genuine non-perturbative phenomena such as poles of the scattering amplitude in the complex plane of the invariant scattering energy $z$, which can be identified with resonances. Coupled channel dynamics plays an essential role in this scheme, with the chiral Lagrangians providing the corresponding interactions of the multiplets; even physically closed channels contribute as intermediate virtual states. 

The $\Lambda(1405)$ and the $N^*(1535)$ have been explained as meson-baryon $(MB)$ quasibound states \cite{Kaiser:1995cy,Kaiser:1996js,Oset:1997it,Nacher:1999vg,Inoue:2001ip,Nieves:2001wt,Oller:2000fj,GarciaRecio:2003ks} from the interaction of the meson octet of the pion (M) with the baryon octet of the nucleon (B), mediated by the Wein\-berg-To\-mo\-za\-wa term. It is interesting to note that, even without chiral Lagrangians, the use of basic interactions for the coupled channels calls for an interpretation of some resonances like the $\Lambda(1405)$ as quasibound states  of the scattering problem \cite{dalitz,Dalitz:1967fp,Jennings:1986yg}. In the last few years, the unitary schemes have been extended to the low lying $3/2^-$ baryonic resonances \cite{Kolomeitsev:2003kt,Sarkar:2004jh} and other quantum numbers \cite{MartinezTorres:2008kh,Gonzalez:2008pv}.

Dynamically generated states can also appear in other approaches.
At higher energies, the dynamics contained in meson exchange is important. Meson exchange models of the $\pi N$ interaction with the analyticity property, developed over the years~\cite{Sato:1996gk,JuliaDiaz:2007kz,Surya:1995ur,Krehl:1999km,Gasparyan:2003fp}, deliver a precise description of the partial waves in $\pi N$ scattering. In Ref. \cite{Krehl:1999km}, the Roper resonance could be described as dynamically generated, i.e., without the need of a genuine (3-quark) resonance state and, in a speed plot, resonance parameters were extracted. In recent studies~\cite{Doring:2009bi,Doring2}, the amplitude could be analytically continued to the complex plane and, indeed, poles of the Roper were found.  

Electromagnetic properties provide additional information about the structure of strongly interacting systems. They allow for independent tests of hadronic models: In the picture of dynamical generation, the mesons and baryons can form a resonance. As the photocouplings to these components are well known, predictions can be made that can be compared to experimental results. In case of the $N^*(1535)$, photoproduction of $\eta$ on $p$ and $n$ has been tested in Ref. \cite{Jido:2007sm}. There, the $Q^2$ dependence of the $A^{1/2}$ and $S^{1/2}$ helicity amplitudes has been evaluated. While there is a qualitative agreement with the data, the faster theoretical fall-off with $Q^2$ indicates a need for additional ingredients in the model such as a small genuine 3-quark state. Similar conclusions have been drawn recently in Ref. \cite{Hyodo:2008xr}.

Many tests of hadronic models with electromagnetic probes have been done at the level of cross sections only (see, e.g. Ref.~\cite{Doring:2006pt}), mainly because of the scarcity of data. 
One can also test the radiative decay of the resonance~\cite{Doring:2006ub,Doring:2007rz}, but its experimental extraction is tied to large uncertainties. A first step towards a test of the amplitude itself and its phase has been done in Ref. \cite{Jido:2007sm}, and the relative signs between $A_{1/2}^p,$ $A_{1/2}^n$ and $S_{1/2}$ could be tested which are in agreement with the results extracted experimentally. Yet, the phase of the photoproduction amplitude is still not accessed through such a study. 

However, a direct access to the photoproduction amplitude is indeed possible via the corresponding multipoles, in this case, $\ezp$. These quantities can be extracted in partial wave analyses that also take into account the polarization observables \cite{Arndt:2002xv,Drechsel:2007if}. This allows to access the relative phases of the $N^*(1535)$ in $\gamma N\to\pi N$ and $\pi N\to\pi N$. As we shall show, this provides a very sensitive test for hadronic models of dynamical generation. In practice, however, one has still to resort to phenomenological analyses which are, in principle, model-dependent because one cannot access the amplitude directly from the existing observables, as there is not yet a ``complete'' experiment available.
Thus, the aim of the present study is to analyze the reactions $\pi N\to\pi N$ and $\gamma N\to\pi N$ at the amplitude level and compare to the phenomenologically extracted multipoles~\cite{Arndt:2002xv,Drechsel:2007if}. 

Observable quantities that
allow one to distinguish between hadronic molecules and more
elementary states are urgently called for. For $S$--wave states close
to thresholds this is discussed in Refs.~\cite{evidence}.
For a model test, a quantitative description of the data is important. This was already stressed in Ref.~\cite{Oller:1998zr,Meissner:1999vr}, where $\pi\pi$ and $\pi N$ scattering
was analyzed within the chiral unitary approach and the
interplay of genuine and dynamically generated resonances was
investigated.

For a quantitative data description of the $S$--wave pion production, the $N^*(1650)$ has to be included. It has the same quantum numbers as the $N^*(1535)$ and lies close-by; the interference among resonances with the same quantum numbers plays an important role as has been recently pointed out in Ref. \cite{Doring2}. The $N^*(1650)$ substantially modifies the position and residue of the $N^*(1535)$. Such an interplay of resonances may even lead to the disappearance of a resonance as seen, e.g., in the $D_{13}$ partial wave discussed in Ref. \cite{Doring2}.

For the dynamical generation of the $N^*(1535)$, we stu\-dy only the $S$-wave pion production as provided by the 
Wein\-berg-To\-mo\-za\-wa term. This effective interaction is clo\-se\-ly tied to $\rho$ meson exchange which, in its full dynamical treatment, contributes to all partial waves and not only to the $S$-wave. Thus, the strong couplings to the $K\Lambda$ and $K\Sigma$ channels, which appear in the description of the $N^*(1535)$ as a dynamically generated resonances, will necessarily have an impact on higher partial waves. While this issue can be studied in the framework of meson exchange models~\cite{Johann_private}, it is beyond the scope of the present $S$-wave model.


\section{Formalism}
\label{sec:forma}


\subsection{Meson-baryon interaction}
\label{sec:hadpart}
Various approaches have been followed in the past to describe the properties of the $N^*(1535)$ in terms of mesons and baryons rather than by the quark degrees of freedom; in Ref. \cite{Kaiser:1995cy,Kaiser:1996js} the chiral meson-baryon Lagrangian through NLO provides the driving interaction. In Ref. \cite{Nieves:2001wt}, an SU(6) fine-splitting in the chiral Lagrangian provides the interaction. We follow here the work of Ref. \cite{Inoue:2001ip} which relies on the use of the leading order Weinberg-Tomozawa term. 

The strength of the meson-baryon interaction is fixed by the pion decay constant $f_\pi$. In $S$-wave projection, the interaction to leading order is given by
\begin{eqnarray}
V^\npo_{i j}(z) &=&
 - C_{i j} \frac{1}{4 f_i\,f_j}(k^0_i+k^0_j)
 \nonumber \\ &&  \ \ \ \times
\sqrt{\frac{M_{i}+E}{2M_{i}}}
\sqrt{\frac{M_{j}+E^{\prime}}{2M_{j}}}
\label{wtt}
\end{eqnarray}
with the channel indices $i,j$, the baryon mass $M$, meson energies $k^0_i$, meson decay constants $f_i$ ($f_\pi,\,f_\eta,\,f_K$), and the center of mass energy $z$. Note the explicit SU(3) breaking from the values of $f_K=1.22\, f_\pi$ and $f_\eta=1.3\,f_\pi$. The value of $f_K$ has been recently evaluated to a higher precision with the slightly changed result of $f_K=1.193\,f_\pi$ \cite{Bernard:2007tk,Amsler:2008zz}. The coefficient $C_{ij}$ is the coupling strength of the meson and baryon, which is determined by the SU(3) group structure of the channel. For strangeness, $S=0$, and isospin,\,$I=1/2$, the values of the coefficient $C_{ij}$ can be found in Refs. \cite{Inoue:2001ip,Doring:2005bx} for the two net charge states $Q=0,\,1$. We work here in the particle basis rather than in the isospin basis which allows to take into account threshold effects from different pion masses in $\pi^0$ photoproduction [cf. Fig. \ref{fig10}]. The six channels for $Q=0,\,1$, respectively, are given in the first column of Table \ref{tab:su3c}. In Eq. (\ref{wtt}), terms of the order $p/M$ have been neglected and we will construct the phototransition amplitudes in agreement with this approximation.

The amplitude in Eq. (\ref{wtt}) is the input for the Bethe-Salpeter equation 
\be
T^\npo &=& V^\npo+V^\npo\,G\,T^\npo\non
       &=&(1-V^\npo\,G)^{-1}V^\npo.
\label{bse}
\ee
The notation $V^\npo$ in Eq. (\ref{wtt}) and $T^\npo$ in Eq. (\ref{bse}) is motivated
by the usual decomposition of the amplitude into the pole ($T^\po$) and non-pole ($T^\npo$) 
parts as defined by Eqs. (\ref{deco1}, \ref{had_deco}) in the following section. 
The potential $V^\npo$ is factorized on-shell as described in detail in Ref. \cite{Oller:1998zr}. Then, the propagator $G$ factorizes and can be explicitly evaluated with the result given, e.g., in Eq. (23) of Ref. \cite{Jido:2007sm}. The factorized propagator $G$ is logarithmically divergent and evaluated with dimensional regularization. For a thorough discussion on the connection of this ansatz to the $N/D$ method \cite{Oller:1998zr} and the relation to the full, non-factorized Bethe-Salpeter equation, see also Ref. \cite{Doring:2006ub,Meissner:1999vr}.

In contrast to the original work on the $N^*(1535)$ \cite{Inoue:2001ip}, in the present study the $\pi\pi N$ channel is not taken into account. The main reason is that a consistent implementation of photoproduction in this channel is beyond the scope of this work; second, the parameterization of the $\pi N\to\pi\pi N$ transition potential from Ref. \cite{Inoue:2001ip} is only valid at lower energies of the energy range considered in the present study. From a practical point of view, the $\pi\pi N$ channel has been found to be small in the isospin $1/2$ channel \cite{Inoue:2001ip} and neglecting it seems to be a valid approximation. We may notice the absence of the multipion states in the isospin $I=3/2$ channel. Also, in $\pi N\to\eta N$, the $\pi\pi N$ channels is necessary to reduce the $\pi N\to\eta N$ cross section from 3 mb to the experimental value of 2.5~mb as pointed out in Ref. \cite{Gasparyan:2003fp}.

It should be pointed out that there are four subtraction constants $a_{\pi N}, \,a_{\eta N},\,a_{K\Lambda},$ and $a_{K\Sigma}$ coming from the logarithmic divergence in the factorized propagator $G$. These constants are regarded as free parameters and used to fit the experimental data. The values of the subtraction constants can be related to the regularization scale \cite{Oller:2000fj} to be fixed at ``natural'' values of around $-2$. In Ref.~\cite{GarciaRecio:2003ks}, approximate crossing symmetry of the amplitude $T$ is imposed by the renormalization condition $T(\sqrt{s}=\mu)=V(\mu)$ and the renormalization scale $\mu$ is related to the baryon masses~\cite{Lutz:2001yb}. A similar ``natural'' renormalization condition has been found recently in Ref.~\cite{Hyodo:2008xr}. 

As strict crossing symmetry is violated at other energies than the renormalization point anyways, we allow for deviations from the natural values of the subtraction constants. Interestingly, it has been pointed out in Ref.~\cite{Hyodo:2008xr} that those deviations can be related to genuine, three-quark like propagator terms in the potential. Thus, as in the present approach such deviations are needed phenomenologically, this is a strong hint that some genuine quark state is necessary for a quantitative description of the $N^*(1535)$ properties in the reactions $\gamma N\to\pi N$ and $\pi N\to\pi N$ studied here. See also Ref.~\cite{Jido:2007sm} where some need for such components has been found in the study of the helicity amplitudes within a similar framework as the present one.

Second, the hadronic interaction in the present approach is mediated by the Weinberg-Tomozawa term, and the pole diagrams (terms with $D$ and $F$) that are present at the same lowest order in the chiral expansion~\cite{Oller:2000fj}, are neglected. In the present approach, we consider contributions from those terms, and also from the contributions of higher order Lagrangians which are not considered here, to be absorbed in the values of the subtraction constants.

In Ref. \cite{Inoue:2001ip}, the four subtraction constants are fitted to the $S_{11}$ and $S_{31}$ $\pi N$ partial wave amplitudes and a pole in $S_{11}$ is found at around $1537-37\,i$ MeV. It should be stressed that the $N^*(1535)$ is dynamically generated even when the natural values for the subtraction constants of Ref. \cite{Hyodo:2008xr} are used. In this case, we find the pole at $z_0=1653-145\,i$ MeV with large couplings into $K\Sigma$ and $\eta N$, in qualitative agreement with the fine tuned model of Ref. \cite{Inoue:2001ip}. Thus, the $N^*(1535)$ and its coupling pattern preexist in the ``natural'' scheme of Ref. \cite{Hyodo:2008xr}, while the fine tuning of the subtraction constants done in Ref. \cite{Inoue:2001ip} merely moves the pole a 100 MeV down in energy into the $N^*(1535)$ region.
Due to its large couplings to $K\Lambda$ and $K\Sigma$, the $N^*(1535)$ is sometimes referred to as a $K\Lambda$, $K\Sigma$ quasibound state.

\subsection{Genuine resonance states}
\label{sec:genureso}
Eq. (\ref{bse}) allows for the formation of poles in the $T$ matrix because the matrix $1-V^\npo G$ can become singular. These poles in the complex plane of the scattering energy $z$ can be identified with resonances on the physical axis (Im $z=0$). In the present approach, this concept of dynamical generation of the $N^*(1535)$ is tested by also allowing for a genuine $N^*(1535)$ via a pole term of the form $1/(z-\bar M_r)$ in the potential. Thus, when performing fits to different data sets, a genuine resonance can replace the dynamical $N^*(1535)$ from the original fit of Ref. \cite{Inoue:2001ip}. This would be a sign that the solution of Ref. \cite{Inoue:2001ip} is unnatural, and that a genuine resonance state in the potential leads to a more natural data description. The criterion to distinguish both scenarios is the quality of the fits to the data. 

A second important point is the presence of the $N^*(1650)$ in the $S_{11}$ partial wave state; it can interfere with the $N^*(1535)$, even far below the $N^*(1650)$ nominal energy of $z=$1.65 GeV \cite{Doring2}. Resonance interference is a necessary ingredient for a realistic description, and thus, we allow in the fits for a second genuine resonance. Both genuine resonances of the form $1/(z-\bar M_r)$
acquire their respective widths through the rescattering via the meson-baryon loops.

For the $S$-wave couplings of the genuine resonance states to the meson-baryon channels $\pi N$, $\eta N$, $K\Lambda$, and $K\Sigma$, we choose a derivative coupling. Such a choice is necessary to fulfill the minimal requirements of chiral symmetry and, in particular, to reduce the influence of resonances at the $\pi N$ threshold. For the derivative coupling, we take the same energy dependence $\sim k^0$ as given in the Weinberg-Tomozawa term from Eq. (\ref{wtt}) itself. In particular, we factorize the $N^*MB$ vertex on-shell in the same way as the Weinberg-Tomozawa term, i.e. $k^0$ is the on-shell energy and not a loop integration variable.

For the SU(3) couplings of the resonances, we assume the resonance to be a superposition of (anti)symmetric octet state $(\bar{8}),\,8$, antidecuplet
$\overline{10}$ and 27-plet. This set of states can be uniquely mapped onto the four physical states $\pi N$, $\eta N$, $K\Lambda$, and $K\Sigma$. 

Altogether, the modification of the interaction kernel from Eq. (\ref{wtt}) by the presence of a genuine resonance is given by the sum of the non-pole part $V^\npo$ from the Weinberg-Tomozawa term and a pole part $V^\po$ from the genuine resonance,
\be
V^\npo_{ij}		&\to& V_{ij} \equiv	V^\npo_{ij}+V^\po_{ij}
=V^\npo_{ij}+\bar{\Gamma}_i \,\bar{D}\,\bar{\Gamma}_j \ ,\non 
\bar{\Gamma}_i	&=&	\frac{k^0_i}{f_\pi}\,\left[g_{8}\,c_s(i)+g_{8'}\,c_a(i)+g_{\overline{10}}\,c_d(i)+g_{27}\,c_{27}(i)\right]\non
\bar{D}^{-1}&=&z-\bar{M}_r,
\label{genterm}
\ee
with the bare resonance masses $\bar{M}_r$ and the bare vertices $\bar{\Gamma}_i$ that depend on the meson energy $k^0$ in channel $i$. $\bar{D}$ stands for the bare propagator. In Eq. (\ref{genterm}), a factor of $1/f_\pi$ has been inserted to have dimensionless couplings $g$. In Table \ref{tab:su3c}, the SU(3) coefficients $c$ are listed; note that the baryon-first coupling scheme is used throughout this study. These coefficients can be obtained from Ref. \cite{McNamee:1964xq}, but in constructing the isospin eigenstates, the minus signs from the phase conventions for $\pi^+$ and $\Sigma^+$ also have to be taken into account. These phases are necessary to ensure that the genuine states are purely isospin $I=1/2$ states. 
The values of $c_s,\,c_a,\,c_d,\,c_{27}$ quoted in Table \ref{tab:su3c} include these additional signs.
\begin{table}
\caption{SU(3) coefficients for the resonance couplings to the meson-baryon channels in the particle basis. Both cases of net charge $Q=0$ and $Q=+1$ are listed. All values carry a square root, i.e. $-3/20$ means $-\sqrt{3/20}$.}
\begin{center}
\begin{tabular}{lllll}
 \hline\hline
$Q=0$		&			&			&			&			\\
Channel $i$	&$\,\,\,\,\,c_s$	&$\,\,\,\,\,c_a$	&$\,\,\,\,\,c_d$	&$\,\,\,\,\,c_{27}$	\\
$K^+\Sigma^-$	&$\,\,\,\,\,3/10$	&$-1/6$			&$\,\,\,\,\,1/6$	&$\,\,\,\,\,1/30$	\\
$K^0\Sigma^0$	&$-3/20$		&$\,\,\,\,\,1/12$	&$-1/12$		&$-1/60$		\\	
$K^0\Lambda$	&$-1/20$		&$-1/4$			&$-1/4$			&$\,\,\,\,\,9/20$	\\	
$\pi^-p$	&$\,\,\,\,\,3/10$	&$\,\,\,\,\,1/6$	&$-1/6$			&$\,\,\,\,\,1/30$	\\	
$\pi^0n$	&$-3/20$		&$-1/12$		&$\,\,\,\,\,1/12$	&$-1/60$		\\	
$\eta n$	&$-1/20$		&$\,\,\,\,\,1/4$	&$\,\,\,\,\,1/4$	&$\,\,\,\,\,9/20$	\\	
$Q=+1$		&			&			&			&			\\	
$\pi^0p$	&$\,\,\,\,\,3/20$	&$\,\,\,\,\,1/12$	&$-1/12$		&$\,\,\,\,\,1/60$	\\	
$\pi^+n$	&$\,\,\,\,\,3/10$	&$\,\,\,\,\,1/6$	&$-1/6$			&$\,\,\,\,\,1/30$	\\	
$\eta p$	&$-1/20$		&$\,\,\,\,\,1/4$	&$\,\,\,\,\,1/4$	&$\,\,\,\,\,9/20$	\\	
$K^+\Sigma^0$	&$\,\,\,\,\,3/20$	&$-1/12$		&$\,\,\,\,\,1/12$	&$\,\,\,\,\,1/60$	\\	
$K^+\Lambda$	&$-1/20$		&$-1/4$			&$-1/4$			&$\,\,\,\,\,9/20$	\\	
$K^0\Sigma^+$	&$\,\,\,\,\,3/10$	&$-1/6$			&$\,\,\,\,\,1/6$	&$\,\,\,\,\,1/30$	\\	
\hline\hline
\end{tabular}
\end{center}
\label{tab:su3c}
\end{table}

\begin{figure}
\begin{center}
\includegraphics[width=0.4\textwidth]{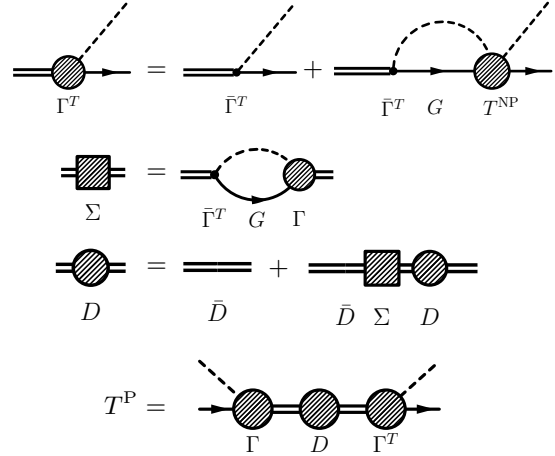}
\end{center}
\caption{Decomposition of the hadronic amplitude in pole and non-pole part according to Eq. (\ref{had_deco}). Diagrammatic representation of bare (dressed) vertices $\bar{\Gamma}$ ($\Gamma$), self energy $\Sigma$, bare (dressed) baryon propagator $\bar{D}$ ($D$) and the pole part $T^\po$.}
\label{fig:reorderhadro}
\end{figure}

It is convenient to decompose the full hadronic amplitude into a non-pole part $T^\npo$ and a pole part $T^\po$ according to (suppressing the channel indices)
\be
T&=&V + VG T \non
 &=&T^P+T^{NP}.
\label{deco1}
\ee
The non-pole part is defined as the sum of those diagrams that do not have a 1-particle cut. This is identical to $T^\npo$ from Eq. (\ref{bse}). A short calculation shows that 
\be
T^\po		&=&{\Gamma}\,\frac{1}{\bar{D}^{-1}-\Sigma}\,{\Gamma}^T,\non 
\Sigma 		&=&{\bar{\Gamma}}^T\,G\,{\Gamma}={\bar{\Gamma}}^T\,(1+\,G\,T^\npo\,G)\,{\bar{\Gamma}},\non 
{\Gamma}	&=&\left(1+T^\npo\,G\right)\,{\bar{\Gamma}}=\left(1-V^\npo\,G\right)^{-1}\,{\bar{\Gamma}} \ ,
\label{had_deco}
\ee
where $\bar{\Gamma}$ and $\Gamma$ are the bare and dressed vertices, respectively;
${\bar{\Gamma}}^T$ indicates the transposed of the vector ${\bar{\Gamma}}$ in channel space. The vertices ${\Gamma}$, self energy $\Sigma$, and the pole part $T^\po$, are displayed in Fig. \ref{fig:reorderhadro}.

In the case of more than one genuine resonance present, Eq. (\ref{had_deco}) becomes
\be
T^\po		&=&\sum_{r,r'} {\Gamma}_{r} \,\frac{1}{\bar{D}_r^{-1} \delta_{rr'} -\Sigma_{rr'}}\,{\Gamma}_{r'}^T,\non 
\Sigma_{rr'} 		&=&{\bar{\Gamma}}^T_r\,G\,{\Gamma}_{r'}={\bar{\Gamma}}^T_r\,(1+\,G\,T^{\rm NP}\,G)\,{\bar{\Gamma}}_{r'}\ ,\non 
{\Gamma}_{r}	&=&\left(1+T^\npo\,G\right)\,{\bar{\Gamma}}_r=\left(1-V^\npo\,G\right)^{-1}\,{\bar{\Gamma}}_r \ ,
\label{had_deco1}
\ee
where the indices $r$ and $r'$ label the resonances; the summation is over the resonances. 

With the formalism of this section, the hadronic part of the model is defined. There are altogether 14 free parameters: 4 subtraction constants for the meson-baryon loops and 4 coupling strengths per genuine resonance state with bare mass $\bar{M}_r$. The model allows for two genuine states and also the formation of dynamically generated poles. In particular, one of the genuine states can replace the dynamically generated $N^*(1535)$ which will be a test of which description of the $N^*(1535)$ is the more appropriate one.


\subsection{Photon couplings to mesons and baryons}
\label{sec:phomeba}
In this section we describe the photon couplings to the meson-baryon components defined in the last section. The photon couplings to the genuine resonance states will be discussed in Sec. \ref{sec:phogen}. They are gauge invariant by themselves and can be discussed separately.

In principle, gauge invariance is ensured by systematically and consistently coupling the photon to all mesons, baryons, and vertices in the rescattering series provided by the Bethe-Salpeter equation. This has been recently discussed \cite{Borasoy:2005zg} and realized in Ref. \cite{Borasoy:2007ku} within the context of $U\chi PT$. In the present case, apart from working in a non-relativistic framework, we have a factorized version of the Bethe-Salpeter equation in Eq. (\ref{bse}), and it is possible to determine subclasses of diagrams whose sum is transversal. 

We summarize at this point a discussion from Ref. \cite{Jido:2007sm} to provide a clear overview of what has been done in the literature \cite{Borasoy:2007ku,Jido:2007sm,Kaiser:1996js} concerning the transversality of the phototransition amplitudes within the $U\chi PT$. A fully consistent treatment has been realized in Ref. \cite{Borasoy:2007ku}. We start the discussion with an approximation to this general scheme, realized in Ref. \cite{Jido:2007sm}, where several electromagnetic properties of the $N^*(1535)$ have been evaluated.
\begin{figure}
\begin{center}
\includegraphics[width=0.43\textwidth]{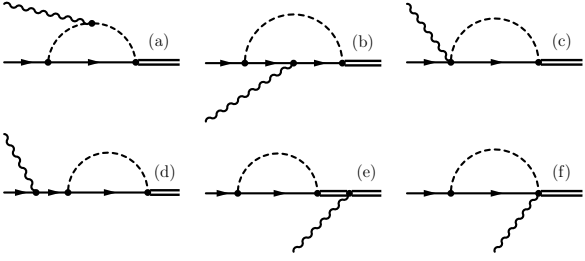}
\end{center}
\caption{Coupling of the photon to loops in $\gamma N\to N^*$. This set is gauge invariant in the fully relativistic treatment. Diagrams (d), (e), and (f) vanish in the heavy baryon limit.}
\label{fig16}
\end{figure}

(I): 
In Fig. \ref{fig16} we  show a set of diagrams with photon couplings in the process $\gamma N\to N^*$ with the quantum numbers of the $N^*$ being $1/2^-$. For the discussion of transversality, we use the relativistic $N$ and $N^*$ propagators and vertices. It has been shown in Ref. \cite{Jido:2007sm} that the set \ref{fig16}(a) to \ref{fig16}(f) is gauge invariant. 

(II): One can factorize the (hadronic) $N^*MB$ vertex in Fig. \ref{fig16} on-shell, i.e., take the $N^*MB$ vertex out of the loop integration. In this case, it has been shown in Ref. \cite{Jido:2007sm} that diagrams (a) to (e) form a transversal set of diagrams. 

(III): A further simplification of the hadronic part of the model is the heavy baryon approach as discussed in Ref. \cite{Jido:2007sm}. In that case, the Dirac structure of the propagators and vertices is simplified. Also in this non-relativistic formalism, the gauge invariance holds order by order in the $1/M$ expansion as discussed in Ref. \cite{Jido:2007sm}. Again, diagrams \ref{fig16}(a) to \ref{fig16}(e) form a transversal set of diagrams. In the non-relativistic treatment, one considers the convection part of the $\gamma(k)B(p)\to B(p')$ coupling, $\sim e(p+p')$,  separately from the magnetic part, $\sim k\times p$; the latter is transverse by itself.

(IV): 
A further simplification has been done, e.g., in Ref. \cite{Kaiser:1996js}. In that case, even the phototransition amplitude itself is factorized on-shell, in addition to the on-shell factorization of the meson-baryon vertices: rather than evaluating the full loops in Fig. \ref{fig16}, an effective range expansion of the tree-level diagrams in Fig. \ref{fig1} is performed. 
The resulting phototransition amplitude is then multiplied with the factorized meson-baryon loop. We will investigate the consequences of such a procedure in Sec. \ref{sec:factoam}. 

The items (I)-(IV) above summarize what has been done in the literature.
In the present study, we evaluate the phototransition amplitude at the level of approximation (III). This means we consider the photon loops (a) to (e) from Fig. \ref{fig16} in the same non-relativistic framework that has been used for the hadronic part of the model discussed in Sec. \ref{sec:hadpart}. In particular, we factorize the on-shell hadronic vertex of the type $MB\to N^*$ on the right-hand sides of the diagrams in Fig. \ref{fig16}. The rescattering scheme from Eq. (\ref{bse}) provides also the Weinberg-Tomozawa term of the type $MB\to MB$ (cf. Fig. \ref{fig:used_pholoops}). This vertex is treated on-shell as well, i.e. it does not contribute in the loop integrations of the loops in Fig. \ref{fig16}. 

This scheme provides, in principle, a well-defined set of diagrams that can be evaluated straightforwardly. It can be realized in a field theoretical approach within an approximation as described in the next section.


\subsection{Formalism of photoproduction}
\label{sec:formaphoto}

In the present investigation, we follow the work of Ref. \cite{Haberzettl:2006bn} to 
construct a photoproduction amplitude based on the field-theoretical approach developed 
by Haberzettl \cite{Haberzettl:1997}. The full formalism of Ref. \cite{Haberzettl:1997} results in a complex and highly non-linear amplitude which requires truncation 
of some of the reaction mechanisms and/or replacement by phenomenological approximations 
for its practical implementation. Any such an approximate treatment should preserve the 
relevant symmetries of the original full amplitude. In particular, unitarity and gauge 
invariance should be maintained. This has been done recently in 
Ref. \cite{Haberzettl:2006bn}, where the complicated part of the interaction current has 
been approximated by a generalized contact current such as to preserve the gauge 
invariance (and unitarity) of the original full amplitude. 

In the following, we show how the treatment of Ref. \cite{Haberzettl:2006bn} is applied 
here. We denote the four-momenta of the initial state photon and nucleon by $k$ and 
$p$, respectively. The four-momenta of the meson and baryon in the final state are
denoted by $q$ and $p'$, respectively. $s\equiv (p+k)^2$, $u\equiv (p'-k)^2$ and 
$t\equiv (q-k)^2$.

In this section we do not consider the genuine resonances; they will 
be included in Sec. \ref{sec:phogen}.

The approximate photoproduction amplitude derived in Ref. \cite{Haberzettl:2006bn} is expressed as
\begin{equation}
M^\mu  = M^\mu_s + M^\mu_u + M^\mu_t + M^\mu_{int} \ ,
\label{eq:d2}
\end{equation}
where $M^\mu_x$ denotes the $x$-channel $(x = s, u, t)$ tree-level amplitude (all involving physical masses and couplings) and $M^\mu_{int}$, the interaction current. The latter is given by
\begin{figure}
\begin{center}
\includegraphics[width=2.cm]{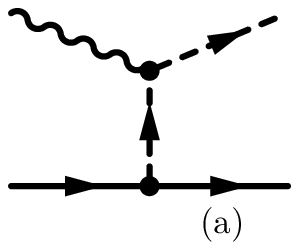}\hspace*{1cm}
\includegraphics[width=2.cm]{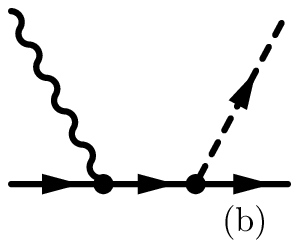}\\ \vspace*{0.3cm}
\includegraphics[width=2.cm]{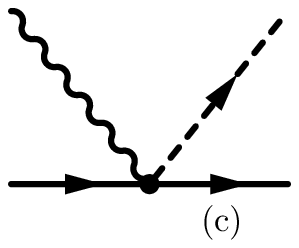}\hspace*{1cm}
\includegraphics[width=2.cm]{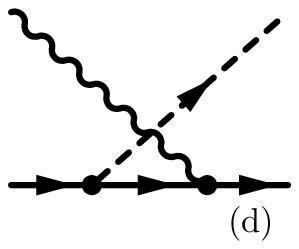}
\end{center}
\caption{Tree-level contributions in photoproduction. Meson pole term (a), baryon pole term (b), Kroll-Ruderman term (c), and crossed nucleon pole term (d). 
They correspond, respectively, to the $t$-channel $M^\mu_t$, $s$-channel $M^\mu_s$, Kroll-Ruderman 
contact $M^\mu_{KR}$, and $u$-channel $M^\mu_u$ currents.}
\label{fig1}
\end{figure}
\begin{equation}
M^\mu_{int}  =  M^\mu_c + T^\mu
  +  T^\npo \tilde{G}\left[ M^\mu_{u T}  + M^\mu_{t T} + T^\mu \right] \ ,  
\label{eq:d9}
\end{equation}
where $M^\mu_{x T}\ (x = u, t)$ stands for the transverse part of $M^\mu_x$, i.e., $k_\mu M^\mu_{x T} = 0$. $T^\mu$ denotes the transverse contact current unconstrained by the Ward-Takahashi identity. As has been mentioned in \cite{Haberzettl:2006bn}, it can be fixed from the data. $M^\mu_c$ (given below) denotes the generalized contact 
current which is an approximation to the complicated part of the interaction current that is not taken into account explicitly. 

\begin{figure}
\begin{center}
\includegraphics[width=0.4\textwidth]{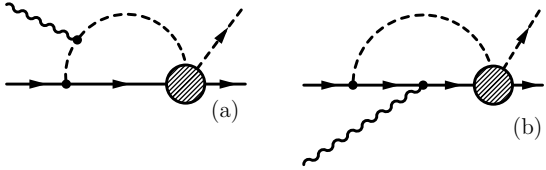}
\end{center}
\caption{Photoproduction loop amplitude [second term in the r.h.s. of Eq.~(\ref{eq:g2})
or Eq.~(\ref{eq:g2cc})]
constructed from $M^\mu_{t T}$, the transverse part of the $t$-channel diagram 
Fig.~\ref{fig1} (a), and from $M^\mu_{u T}$,the transverse part of the $u$-channel 
diagram Fig.~\ref{fig1} (d). The intermediate loops imply a sum over allowed 
SU(3) states. The shaded circle indicates the factorized hadronic amplitude, $T^{NP}$
from Eq. (\ref{bse}).}
\label{fig:used_pholoops}
\end{figure}

The terms with $\tilde{G}$ in Eq. (\ref{eq:d9}) imply an integration over the loop four-momentum, which is not indicated explicitly. In particular, the loop function is not factorizing. In order to distinguish these terms from the previous formalism of the hadronic amplitude, where $G$ is indeed factorizing, we denote the non-factorizing terms with $\tilde{G}$ instead of $G$. Furthermore, throughout this paper, we refer to this loop ($\tilde G$) as the {\it photon loop} since it involves a photon attaching to one of the particles (meson or baryon) in the loop integral as illustrated diagrammatically in Fig.~\ref{fig:used_pholoops}. This is not to be confused with the intermediate $\gamma B$ loop which would be present in a coupled channel formalism beyond one-photon approximation. The present formalism is within an one-photon approximation.

Unlike some meson exchange models, the pre\-sent model contains no (phenomenological)
form factors. In this case, 
the generalized contact current $M^\mu_c$, as prescribed in Ref. \cite{Haberzettl:2006bn}, reduces to the usual 
(dressed) Kroll-Ruderman contact current
\begin{equation}
M^\mu_c = M^\mu_{KR} \equiv  \Gamma^\mu_{BBM\gamma} \ ,
\label{eq:g1}
\end{equation}
where $\Gamma^\mu_{BBM\gamma}$ denotes the $BBM\gamma$ Kroll-Ruderman contact vertex.

As mentioned before, the transverse contact current $T^\mu$, unconstrained
by gauge invariance, can, in principle, be fixed from the data. In the present
work, however, we set it to $T^\mu = 0$ for simplicity; the existing data
show no clear sign for its necessity.
(See, however, the discussion of the $E_{0+}$ multipole amplitude for neutral pion production in 
Sec.~\ref{sec:fit1}.)

The approximate amplitude [cf. Eqs.~(\ref{eq:d2},\ref{eq:d9})] then can 
be written as
\begin{equation}
M^\mu  =  V^\mu + T^\npo \tilde{G}\left[M^\mu_{u T} + M^\mu_{t T}\right] \ , 
\label{eq:g2}
\end{equation}
where 
\begin{equation}
V^\mu \equiv  M^\mu_s + M^\mu_u + M^\mu_t + M^\mu_{KR}
\label{eq:g2a}
\end{equation}
constitutes the sum of the usual tree-level Feynman diagrams which is gauge invariant. 
It is illustrated diagrammatically in Fig. \ref{fig1}. The second term in the r.h.s. 
of Eq.~(\ref{eq:g2}) is illustrated in Fig. \ref{fig:used_pholoops}. 

The above equation defines the photoproduction amplitude of the present model excluding
the genuine resonances. 
Note that, in general, a regularization procedure is required in order to carry 
out the integral in Eq. (\ref{eq:g2}). In the remaining of this section, we show 
how this is done here. Using the notation
\begin{equation}
\tilde M \equiv \braket{\bar u(\vec p\ ')| M^\mu\epsilon_\mu |u(\vec p)} \ , 
\label{eq:g3}
\end{equation}
where $\ket{u(\vec p)}$ stands for the nucleon spinor without the Pauli spinor 
and the normalization $\braket{\bar u(\vec p)|u(\vec p)} = 1$, 
Eq.~(\ref{eq:g2}) becomes,
\begin{equation}
\tilde M  =   \tilde V  + \tilde{M}_{loop} \ ,
\label{eq:g4}
\end{equation}
where 
\begin{equation}
\tilde M_{loop}  \equiv \bra{\bar u(\vec p\,')}
T^\npo \tilde{G}\left[M^\mu_{u T} + M^\mu_{t T}\right] \epsilon_\mu \ket{u(\vec p)}\ .
\label{eq:g5}
\end{equation}
In the present work, we calculate the integral in the above equation in the 
non-relativistic limit ($|\vec p| \ll M$). Also, our model uses an on-shell factorized 
final state interaction $T^\npo$. With these approximations, it can be shown~\cite{Marco:1999df,Jido:2007sm,Doring:2007rz} that the loop amplitude $\tilde M_{loop}$ in the above 
equation decomposes into the form (in the c.m. frame of the system) 
\begin{equation}
\tilde M_{loop}  =  \epsilon_\mu \Gamma^{\mu\nu} \sigma_\nu \ ,
\label{eq:g6}
\end{equation}
where $\sigma_\nu \equiv (0, \vec\sigma)$ and
\begin{equation}
\Gamma^{\mu\nu} \equiv a\,g^{\mu\nu} + b\,P^\mu P^\nu + c\,k^\mu P^\nu 
                   + d\,P^\mu k^\nu + e\,k^\mu k^\nu \ , 
\label{eq:g7}
\end{equation}
with $P \equiv p + k = q + p' = (P^0=z,\; \vec 0)$.

By construction, $\Gamma^{\mu\nu}$ is transverse, i.e., $k_\mu\Gamma^{\mu\nu}\sigma_\nu = 0$ 
since $\tilde M_{loop}$ is transverse. This leads to the relation among the 
coefficients in Eq.~(\ref{eq:g7})
\begin{equation}
a + d\,k\cdot P + e\,k^2 = 0 \ .
\label{eq:g8}
\end{equation}
Note that in Eq.~(\ref{eq:g7}) the only terms that actually contribute to the 
loop amplitude in Eq.~(\ref{eq:g6}) are the $a$ and $d$ terms. The other terms 
contribute nothing once they are contracted with $\sigma_\nu$ and the photon 
polarization vector $\epsilon_\mu$. Therefore, within the present approximation, 
the loop amplitude has the structure 
\begin{equation}
\tilde M_{loop} = - a\;\vec{\sigma}\cdot\vec{\epsilon} 
                  - d\;\vec{\sigma}\cdot\vec{k}\;\epsilon_0\,P^0 \ .
\label{eq:g9}
\end{equation}
For photoproduction, only the $a$ term contributes in the above equation, while  
both the $a$ and $d$ terms contribute to electroproduction. In the former reaction, 
from Eq.~(\ref{eq:g8}), the coefficient $a$ is related to coefficient $d$ by
\begin{equation}
a = - d\,k\cdot P \ .
\label{eq:g10}
\end{equation}
It should be noted that although the $e$ term in Eq.~(\ref{eq:g7}) does not 
contribute to the loop amplitude $\tilde M_{loop}$ [cf. Eq.~(\ref{eq:g9})],
it is crucial for the proper gauge invariance condition given by Eq.~(\ref{eq:g8}) 
which is relevant for electroproduction. 

It is instructive to compare the structure of the loop amplitude given
by Eq. (\ref{eq:g9}) with the most general form of the electroproduction 
amplitude \cite{Berends:1967vi},
\begin{eqnarray}
i\tilde M_{full} & = & i J_1 \vec\sigma\cdot\vec\epsilon
    +   J_2 \vec\sigma\cdot\hat q \vec\sigma\cdot (\hat k \times \vec \epsilon)
    + i J_3 \vec\sigma\cdot\hat k \hat q\cdot\vec\epsilon \non
& + & i J_4 \vec\sigma\cdot\hat q \hat q\cdot\vec\epsilon 
    - i J_7 \vec\sigma\cdot\hat q \epsilon_0 
    - i J_8 \vec\sigma\cdot\hat k \epsilon_0 \ .
\label{Jampl-S}
\end{eqnarray}
A comparison of Eq.~(\ref{eq:g9}) with the above equation shows that the 
$a$ term corresponds to the $J_1$ term and the $d$ term to $J_8$. All other 
structures $J_2$ to $J_7$ are zero in the loop amplitude Eq.~(\ref{eq:g9}).
It is also immediate that the $S$-wave state can only contribute to the $J_1$ 
and $J_8$ terms in the above equation ($a$ and $d$ terms in Eq.~(\ref{eq:g9})).  

As mentioned before, our model treats the loop integral in Eq.~(\ref{eq:g5}) in 
the non-relativistic limit, in addition to using an on-shell factorized $T^\npo$. 
With these approximations, the coefficient $d$ in Eq.~(\ref{eq:g9}) can be extracted 
straightforwardly by a direct integration of the loop in Eq.~(\ref{eq:g5}) rendering 
a well defined (finite) value, a feature that can be shown from dimensional 
considerations. On the other hand, the direct loop integration yields infinity for the 
coefficient $a$ which, as mentioned before, calls for a regularization of the loop 
integral. We do this effectively by determining $a$ via Eq.~(\ref{eq:g10}). In the 
following subsection, some details on the evaluation of the coefficient $d$ is given. 

Eq.~(\ref{eq:g2}) can be extended to a coupled-channel approach as
\begin{equation}
M^\mu_i  =  V^\mu_i + \sum_j T^{NP}_{ij}\tilde{G}_j\left[M^\mu_{u T} + M^\mu_{t T}\right]_j \ , 
\label{eq:g2cc}
\end{equation}
where the subscript $i(j)$ specifies the baryon-meson channel.

Finally, before leaving this section, we mention that the present photoproduction amplitude 
given by Eqs. (\ref{eq:g2}, \ref{eq:g2a}) can also be obtained from the full amplitude given by 
Eq.~(3.20) of Ref.~\cite{Borasoy:2007ku}, provided we identify the term $M^\mu_a$ of 
Ref.~\cite{Haberzettl:2006bn} --- which is to be approximated by a generalized contact current ---
with the sum of the bare Kroll-Ruderman term (the first diagram in Fig.~3(F) ) and diagram 
Fig.~3(G) of Ref.~\cite{Borasoy:2007ku}. In addition, the dressed photon vertex of 
Ref.~\cite{Haberzettl:2006bn} should be 
identified with the dressed photon vertex of Ref.~\cite{Borasoy:2007ku}, represented there in Fig.~3
by the photon line attached to the open square.

  
\subsubsection{Evaluation of the photon loops}
\label{sec:photoprod}
In this section we give some details on the calculation of the coefficient $d$ appearing 
in Eqs. (\ref{eq:g7}, \ref{eq:g9}). 
In the following, since we are interested only in the $\gamma N \to N\pi$ process 
in the present work, we restrict ourself to this channel, i.e., $i=N\pi$ in 
Eq.~(\ref{eq:g2cc}). 

With the on-shell factorization of the FSI, $T^\npo$, the amplitude from Eq. (\ref{eq:g9}) can be written as ($i=N\pi$)
\be
i\,\tilde{{\cal M}}_{loop}&=&\sum_j T^{NP}_{ij}(z) \,\tilde\Gamma_j \ ,\non
\tilde\Gamma_j & \equiv & \int d^4p''\, \left[\tilde{M}_{u T}(p'') + \tilde{M}_{t T}(p'')\right]_j 
\tilde G_j(p'') \ , \ \ \ \ \non
\label{tfact}
\ee
where we have displayed only the relevant argument on which each quantity depends 
upon. $\tilde\Gamma_j$ can be expressed in the form
\be
\tilde\Gamma_j &=& -\frac{i}{2f_\pi}\,A_j\,k\cdot P\,\tilde{d}_j\,\vec{\sigma}\cdot\vec{\epsilon},\non 
k\cdot P&=&\frac{z^2-M_N^2}{2}
\label{vgnonpole}
\ee
where factors $A_j$ for channel $j$ combine the SU(3) factors of the $MBB$ and the $\gamma MM$ 
vertices. They are given in Table \ref{tab:vganp} for each channel. $\tilde d_j$ is the
part of the loop integral contributing to the coefficient $d$ in Eq.~(\ref{eq:g7})  
expressed in terms of the Feynman parameter integrals. The contribution to it from diagram 
\ref{fig:used_pholoops}(a) is finite with the result
\be
&&\tilde d_j^{\,\ref{fig:used_pholoops}(a)} =-|e_j|
\frac{4\,M_j}{(4\pi)^2}\int\limits_0^1 dx\int\limits_0^{1-x}dz\non
&&\times\,\frac{x(z-1)}{x[(x-1)s+z(s-M_N^2)+M_j^2]+(1-x)m_j^2},\non
\label{dterma}
\ee
where $M_j\,(m_j)$ is the baryon (meson) mass in the loop, $M_N$ is the mass of the incoming nucleon (proton or neutron). The factor $|e_j|$ is the modulus of the charge of the loop meson (0 or $+e$). 

\begin{table}
\caption{Factors $A_j$ for the photon loops in the different charge channels $j$. For the channels $K^0\Sigma^0$, $K^0\Lambda$, $\pi^0n$, and $\eta n$, $A_j=0$.}
\begin{center}
\begin{tabular}{llll}
 \hline\hline
$\pi^0p$		&$\pi^+n$		&$\eta p$		&$K^+\Sigma^0$		\\
-$(D+F)$		&$\sqrt{2}(D+F)$	&-$(3F-D)/\sqrt{3}$	&$D-F$			\\
$K^+\Lambda$		&$K^0\Sigma^+$		&$K^+\Sigma^-$		&$\pi^-p$		\\
-$(D+3F)/\sqrt{3}$	&-$\sqrt{2}(D-F)$	&$\sqrt{2}(D-F)$	&-$\sqrt{2}(D+F)$	\\
\hline\hline
\end{tabular}
\end{center}
\label{tab:vganp}
\end{table}

The contribution to $\tilde d_j$ from diagram \ref{fig:used_pholoops}(b) is subleading in $1/M$. Yet, it is evaluated for a discussion on theoretical uncertainties in Sec. \ref{sec:fit2} with the result
\be
& &\tilde d_j^{\,\ref{fig:used_pholoops}(b)} = |e_{B_j}|\frac{4\,M_j}{(4\pi)^2} \int\limits_0^1 dx\int\limits_0^{1-x}dz \non
& &\times \frac{x\,z}{x[(x-1)s+z(s-M_N^2)+m_j^2]+(1-x)M_j^2}, \non
\label{dtermb}
\ee
where $|e_{B_j}|$ is the modulus of the charge of the loop baryon. We have used here only the convection part of the $\gamma BB$ coupling as discussed in the following.

Diagram \ref{fig:used_pholoops}(b) contributes to the $d$ term but the contribution is subleading \cite{Doring:2005bx,Doring:2007rz} in $1/M$. We neglect this term for the numerical results, but consider it in the discussion on theoretical uncertainties in Sec. \ref{sec:fit2}. 

There are various contributions to diagram \ref{fig:used_pholoops}(b) as discussed in detail in Ref. \cite{Jido:2007sm}. Apart from the convection part of the $\gamma BB$ coupling, evaluated in Eq. (\ref{dtermb}), there is a magnetic part. Both parts are of similar and of small sizes \cite{Jido:2007sm}. Furthermore, in Ref. \cite{Jido:2007sm} the $\Sigma^0\Lambda$ transition magnetic moment has been considered, whose contribution also results to be small. In particular, all these different contributions for diagram \ref{fig:used_pholoops}(b) are not only small in size, but barely change the phase of the phototransition amplitude (cf. Table V of Ref. \cite{Jido:2007sm}).


\subsubsection{Photocouplings to genuine resonance states}
\label{sec:phogen}
Apart from the phototransitions via loops, we also allow for direct $\gamma NN^*$ coupling to the genuine resonance states.
For the photon couplings to a genuine $N^*=S_{11}$ resonance we consider the Lagrangian
\be
{\cal L}_{N^*N\gamma}=\bar{N}^*\,\gamma_5\,\frac{g_{\gamma NN^*}}{2\,M_N}\,\sigma_{\mu\nu}\,\partial^\nu\,A^\mu\,N 
\label{fullphogen}
\ee
which provides a pure transversal $\gamma N\to N^*$ transition. The transversality holds order by order in momentum, which can be also shown explicitly. In the non-relativistic reduction that is used in this study, Eq. (\ref{fullphogen}) leads to the vertex
 \be
\epsilon_\mu \Gamma^\mu_{N^*N\gamma} = \frac{\,g_{\gamma NN^*}\,k^0\,(\vec{\sigma}\cdot\vec{\epsilon})}{2\,M_N}+{\cal O}\left(\frac{k}{M}\right)
 \label{nonre}
 \ee
where $k$ is the momentum of the incoming photon and $g_{\gamma NN^*}$ is a free parameter. In the present model, we allow for two genuine resonances, which both appear in charge zero and charge +1. Thus, the vertex from Eq. (\ref{nonre}) appears four times in the model, with four independent coupling constants $g_{\gamma pN^*}^{(1,2)}$, $g_{\gamma nN^*}^{(1,2)}$.

\begin{figure}
\includegraphics[width=0.34\textwidth]{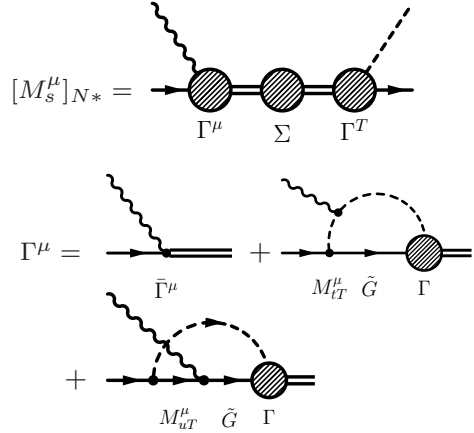}
\caption{Photoproduction amplitude with $s$-channel genuine resonances, corresponding to Eq.~(\ref{eq:c8c}). $\bar{\Gamma}^\mu$ ($\Gamma^\mu$) is the bare (dressed) photon vertex [Eq.~(\ref{eq:d10c})]. $\Gamma$ ($\Sigma$) is the dressed hadronic vertex (self energy).}
\label{fig:gengam}
\end{figure}

The inclusion of genuine resonance states leads to an additional contribution to the photoproduction amplitude given by Eq.~(\ref{eq:d2}). This additional contribution is given by
\begin{equation}
\left[M^\mu_s\right]_{N^*} = \sum_{rr'} {\Gamma}_{r}\,\frac{1}{\bar{D}_r^{-1}\delta_{rr'}-\Sigma_{rr'}}\,
\Gamma^\mu_{r'}\ ,
\label{eq:c8c}
\end{equation}
where the summation runs over the resonances labelled by indices $r$ and $r'$. In the present study,
the $N^*N\gamma$ electromagnetic transition vertex, 
$\Gamma^\mu_{r'}$, in the above equation is dressed according to
\begin{equation}
{\Gamma^\mu}_{r'}  =  \bar{\Gamma}^\mu_{r'} 
 + \left[M^\mu_{u T} + M^\mu_{t T}\right]\,\tilde{G}\,\Gamma_{r'}
\label{eq:d10c}
\end{equation}
for a given resonance labelled $r'$. A diagrammatic representation of Eqs.~(\ref{eq:c8c}, \ref{eq:d10c}) is shown in Fig. \ref{fig:gengam}.
In the actual calculation, the $\gamma NN^*$ vertex from Eq. (\ref{nonre}) is factorized on-shell.
Note that, in principle, the genuine resonances can also contribute to the $u$-channel amplitude $M^\mu_u$. We have ignored this contribution in the present work.

With the inclusion of the genuine resonances as described above, the full photoproduction amplitude in the present work becomes
\begin{equation}
\tilde M_{full} = \tilde M + [\tilde M_s]_{N^*} \, ,
\label{eq:g4c}
\end{equation}
where $\tilde M$ is given by Eq.~(\ref{eq:g4}) and 
\[ 
[\tilde M_s]_{N^*} \equiv \braket{\bar u(\vec p\,')| \left[M^\mu_s\right]_{N^*}\epsilon_\mu |u(\vec p)}.
\] 

\subsubsection{Tree--level amplitudes}
\label{sec:treelevel}
Since in the present work the hadronic processes are described in the $S$ partial-wave state [cf. 
Secs. \ref{sec:hadpart}, \ref{sec:genureso}], the nucleon $s$-channel amplitudes 
$M^\mu_s$ won't be dressed by the hadronic interaction in the $P$-wave state. Furthermore, since 
the loop integrals are evaluated in the non-relativistic limit, there are no dressings of the 
nucleon amplitude $M^\mu_s$ in the $S$-wave (negative energy) state either. Therefore, 
we evaluate the nucleon contribution to the tree-level amplitude $\tilde V$ 
in Eq.~(\ref{eq:g4}) [cf. diagrams in Fig.~\ref{fig1})], including the $s$-channel amplitude 
$M^\mu_s$, using the physical nucleon mass and physical coupling constants. 
This is also in accordance with the gauge invariant approach of Ref.~\cite{Borasoy:2007ku} , where
the photon-baryon vertex (represented by the photon line attached to open square in Fig.~3 in that reference) 
is not dressed by the equation.
In contrast, the genuine 
$S_{11}$ resonances are dressed as mentioned in the previous section. 

Also, the nucleon tree-level amplitudes are evaluated with the full Dirac structure and relativistic propagators and then projected onto $S$-wave to filter the $S$-wave contribution to photoproduction we are interested in. 

The photoproduction amplitude $V^\mu$ is constructed from the following 
Lagrangian densities
\begin{subequations}
\begin{align}
{\cal L}_{NN\pi} & = - \frac{D+F}{2\,f_\pi} \bar N \left( \gamma_5\, 
\fs{\partial} \vec{\pi}\cdot\vec\tau \right) N ~,\\[1ex]
 {\cal L}_{NN\gamma}   &= - e\, \bar N\left( \left[\hat e \gamma^\mu -
\frac{\hat\kappa}{2M_N} \sigma^{\mu\nu}\partial_\nu \right] A_\mu \right)\, N ~,\\[1ex]
 {\cal L}_{NN\pi\gamma}  &=  e\, \frac{D+F}{2\,f_\pi} \,
 \bar N \gamma_5\fs{A} [\vec\tau \times \vec\pi]_3\, N  ~,\\[1ex]
 {\cal L}_{\pi\pi\gamma}  &= e\, [(\partial^\mu \vec\pi) \times \vec\pi]_3 A_\mu ~.
\end{align}
\end{subequations}
Note that in the above Lagrangian we have also taken into account the anomalous term, which is higher order in photon momentum.
Here, $N$ and $\vec{\pi}$ denote the nucleon and pion fields, respectively. The 
vector notation refers to the isospin space. We take $D+F=1.37$. $A_\mu$ denotes the electromagnetic field. $e$ is the 
proton charge; $\hat e = (1 + \tau_z)/2$ and $\hat\kappa = [1.79(1+\tau_z)/2 - 
1.93(1-\tau_z)/2]$ are the charge and anomalous magnetic moment operators of the 
nucleon, respectively. 

The propagators required for constructing the tree-level amplitudes are
\begin{equation}
 \Delta(q) = \frac{1}{ q^2 - m_\pi^2 }  ~,\ \ \ \ \ \ \ \ 
 S(p) =  \frac{1}{\fs{p}-M_N}  ~,
\label{propags}
\end{equation}
where $\Delta(q)$ denotes the pion propagator with mass $m_\pi$ and, 
$S(p)$, the nucleon propagator with mass $M_N$.

\subsection{Observables and analysis}

Summarizing the present model, we have four free parameters from the subtraction constants of the channels $\pi N,\, \eta N, \,K\Lambda, \,K\Sigma$ (6 channels in the particle basis). For the two genuine resonance states allowed in the model we have, for each of them, four couplings to meson-baryon plus two couplings to the photon as free parameters, in addition to the bare mass. Altogether, we thus have 18 free parameters.

In order to compare to cross sections and/or partial wave analyses, we quote their connections with $T$ and $\tilde M_{full}$ as defined in this study. For $\pi N\to MB$, the hadronic amplitude $T$ given in Sec.~\ref{sec:genureso} is connected to the dimensionless partial wave amplitude ${\tilde T}$ from Ref. \cite{Arndt:2008zz} by
\be
{\tilde T}_{\pi N\to MB}&=&-\sqrt{\rho_{\pi N}\,\rho_{MB}}\,T_{\pi N\to MB},\non
\rho_{MB}&=&\frac{M_B\,Q_{MB}}{4\pi\, z}
\label{conn}
\ee
where $M_B$ is the mass of the outgoing baryon and $Q_{MB}$ is the magnitude of the c.m. three-momentum in the outgoing MB channel. The total cross sections for $\pi N\to MB$ are
\be
\sigma_{\pi N\to MB} =\frac{1}{4\pi}\,\frac{Q_{MB}}{Q_{\pi N}}\,\frac{M_N\,M_B}{s}\,|T_{\pi N\to MB}|^2 .
\ee
Similarly, for the $\gamma(k) N\to MB$ process, the total cross sections are given by
\be
\sigma_{\gamma N\to MB}&=&\frac{1}{4\pi}\,\frac{M_N\,M_B}{s}\,\frac{Q_{MB}}{|\vec k|}\, \non
&\times&\frac{1}{4}\sum_{m'm\lambda} \left| \braket{\frac{1}{2} m'|\tilde M^\lambda_{\gamma N\to MB}
|\frac{1}{2} m}\right|^2 \ ,
\ee
where $\tilde M^\lambda_{\gamma N\to MB}$ stands for the photoproduction
amplitude, $\tilde M_{full}$, given by Eq.~(\ref{eq:g4c}). $\lambda$ stands for the two independent 
polarizations of the photon beam. $m'(m)$ denotes the spin-projection quantum number of the nucleon 
in the final(initial) state.

Since we restrict ourselves to the $S$ partial-wave only, the photoproduction amplitude 
$\tilde M_{full}$ is related to the familiar multipole amplitude $E_{0+}$ \cite{Berends:1967vi}
by
\be
\ezp=\frac{\sqrt{M_N\, M_B}}{4\pi\,z}\, J_1.
\label{ezp}
\ee
where $J_1$ is from Eq.~(\ref{Jampl-S}).

In terms of $E_{0+}$, the photoproduction total cross sections are
\be
\sigma_{\gamma N \to BM}= 4\pi\,\frac{Q_{MB}}{|\vec k|}\,|\ezp|^2_{\gamma N \to MB} \ .
\ee

The $\ezp$ multipole in the isospin basis for the outgoing particles is given by
\begin{align}
_p\ezp (S_{11})& =\frac{\sqrt{2}}{3}\,\ezp (n\pi^+)	 +\frac{1}{3}\,\ezp (p\pi^0),\non 
_n\ezp (S_{11})& =\frac{\sqrt{2}}{3}\,\ezp (p\pi^-)	 -\frac{1}{3}\,\ezp (n\pi^0),\non 
\ezp (S_{31})  & =\ezp (p\pi^0)				 -\frac{1}{\sqrt{2}}\,\ezp (p\pi^+) ,
\label{isospinezp}
\end{align}
that show an additional factor of $1/\sqrt{3}$ compared to Clebsch-Gordan coefficients.

The $T$ and $\tilde M_{full}$ amplitudes have poles in the complex plane of the scattering energy $z$. These poles lie on the unphysical sheets. It is thus, necessary to analytically continue the amplitude to these sheets. This is a standard procedure and described in detail in Sec. \ref{sec:analcont}. Poles of $T$ or $\tilde M_{full}$ in the complex plane can come from the genuine resonance states in the model, but also from the unitarization via the Bethe-Salpeter equation (\ref{bse}); indeed, the non-pole part from Eq. (\ref{deco1}) itself obeys the Bethe-Salpeter equation. In particular, the term $1-V^{\rm NP} G^{(2)}$ can become singular if the Weinberg-Tomozawa term $V^{\rm NP}$ provides sufficient attraction. This will lead to a ``dynamically generated'' pole in $T$. 

To analyze the poles, it is convenient to perform a Laurent expansion at the pole position at complex $z_0$. The leading term provides the residue $a_{-1}$ and the hadronic amplitude can be written as
\be
T^{ij}&=&\frac{a_{-1}}{z-z_0}+a_0+a_1(z-z_0)+{\cal O}(z^2),\non 
a_{-1}&\equiv & g_i\,g_j
\label{pa}
\ee
for a transition from channel $i$ to $j$. The residues are parameterized as products of values $g_i\,g_j$ which we call coupling strengths. We call Eq. (\ref{pa}), through the residue term with $a_{-1}$, the {\it pole approximation} (PA) of the amplitude.


\section{Results}
\label{sec:results}
\subsection{The $\ezp$ multipole and the phase problem}
\label{sec:phasprob}
In this section, we will see that there is a serious phase problem in $\ezp$ tied to the model of Ref. \cite{Inoue:2001ip}. We will also see that in order to resolve the issue, we have to allow for the presence of additional resonances.

Before coming to the results of the present model introduced in Sec. \ref{sec:hadpart}, we show the $\ezp$ multipole evaluated from the original model of Ref. \cite{Inoue:2001ip}. In that model, no genuine resonances are present and the $N^*(1535)$ is fully dynamically generated from the coupled channel interaction and the unitarization from Eq. (\ref{bse}). This means, the interaction kernel $V$ is entirely given by the Weinberg-Tomozawa term $V^\npo$ from Eq. (\ref{wtt}). As there are no bare $\gamma NN^*$ couplings, the photon interaction is given entirely by $\tilde{M}$ from Eq. (\ref{eq:g4}).

\begin{figure}
\includegraphics[width=0.43\textwidth]{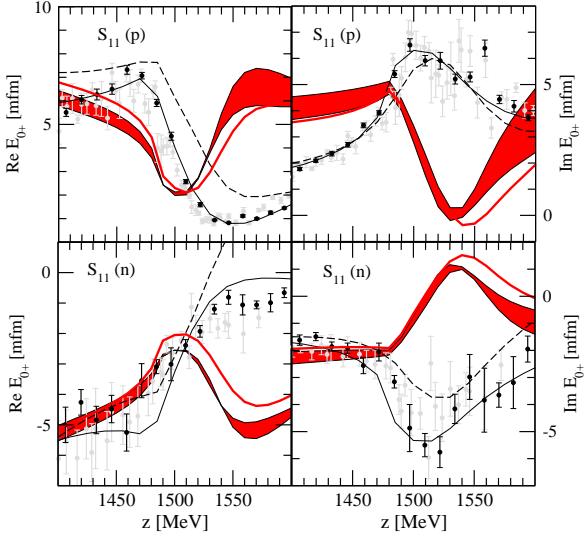}
\caption{Multipoles $\ezp$ in $S_{11}$. The red band shows the prediction using the full and simplified model from  Ref. \cite{Inoue:2001ip}. The solid red line shows the prediction of a refit to $\pi N$ scattering within the full model of Ref. \cite{Inoue:2001ip}. The theoretical curves have been shifted upwards/downwards to have a better overlay with the data. Black data points: SES $\gamma N\to \pi N$ partial wave analysis from Ref. \cite{Arndt:2008zz} [FA07]. Gray points: SES from Ref. \cite{Arndt:2002xv} [FA06]. Solid (dashed) black lines: Analyses SAID from \cite{Arndt:2008zz} [FA07] (MAID2007 from \cite{Drechsel:2007if}).}
\label{fig:oriprob}
\end{figure}

The results for $_p\ezp$ (photoproduction on the proton) and $_n\ezp$ (neutron), using the model from Ref. \cite{Inoue:2001ip}, are shown in Fig. \ref{fig:oriprob}.
To compare the energy dependence, we have slightly shifted the theoretical results upwards/ downwards to match them to the experimental $\ezp$ (in the following, we ignore this discrepancy and concentrate on the resonance shape). The red band in Fig. \ref{fig:oriprob} is given by two different solutions (full and simplified) from Ref. \cite{Inoue:2001ip}: in the simplified version of the original full model the $\pi\pi N$ channel and the form factors from the Weinberg-Tomozawa term are omitted \cite{Inoue:2001ip}. Yet, as Fig. \ref{fig:oriprob} shows, the outcome is not very sensitive to these details. 

In the model of Ref.~\cite{Inoue:2001ip}, there are residual discrepancies with the $\pi N$ phase shifts. The deviations observed for $\ezp$ in Fig. \ref{fig:oriprob} may be due to these discrepancies. Thus, we have performed a refit to $\pi N\to\pi N$ using the full model from Ref. \cite{Inoue:2001ip}, but choosing only a narrow energy interval around the $N^*(1535)$ and not fitting to the $S_{31}$ partial wave amplitude. The fit parameters are the four subtraction constants $a_i$ for the loop functions of the channels $\pi N$, $\eta N$, $K\Lambda$, and $K\Sigma$. In the refit the theoretical solution in $\pi N\to\pi N$ matches much better the results of the $\pi N$ PWA analyses \cite{Arndt:2008zz,Arndt:2002xv} in the $N^*(1535)$ region. In particular, in the refit the $N^*(1535)$ becomes wider while the fitted subtraction constants are still close to their original values quoted in Ref.~\cite{Inoue:2001ip}. Yet, when evaluating $\ezp$ from this refit, the result stays qualitatively the same, indicated with the red solid line in Fig. \ref{fig:oriprob}.

Thus, the $N^*(1535)$ from the model of Ref. \cite{Inoue:2001ip} is seriously off phase in $\ezp$, and this result does not depend on the details of that model, nor can it be easily cured with a refit, which delivers a better agreement in $\pi N\to\pi N$. Note that the Watson's theorem relates the phase of the pion photoproduction amplitude with the $\pi N$ phase-shifts for energies up to the first open channel. For energies above this opening, this theorem doesn't hold anymore. The energy region of interest here is in the second resonance energy region where the $\pi\pi N$ and $\eta N$ channels are opened \footnote{Strictly speaking, the opening of the $\pi^- p$ channel which is just about 5 MeV above the $\pi^0 p$ threshold already invalidates the applicability of the Watson's theorem.}.

\begin{figure}
\includegraphics[width=0.48\textwidth]{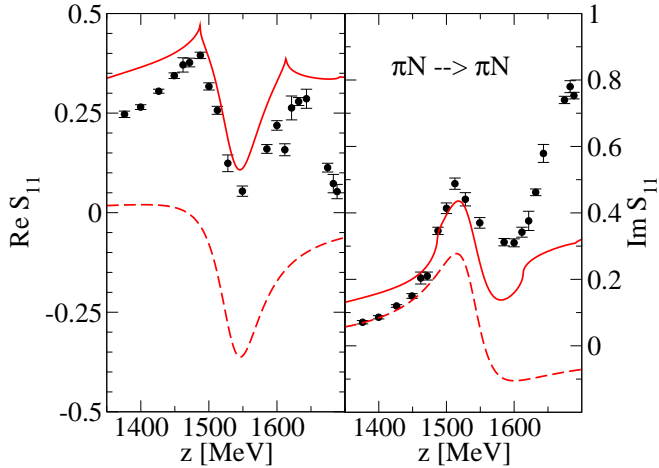}
\caption{Simplified version without $\pi\pi N$/ form factors of the original model from Ref. \cite{Inoue:2001ip} in $\pi N\to\pi N$ (red solid line). The full model from Ref. \cite{Inoue:2001ip} shows a slightly better agreement with the $\pi N$ PWA. The red dashed line shows the pole approximation from Eq. (\ref{pa}). The data points are from the PWA of Ref. \cite{Arndt:2003if}.}
\label{fig:oris11}
\end{figure}

In order to understand the origin of this problem, we consider the pole approximation from Eq. (\ref{pa}) for $\pi N\to\pi N$. In Fig. \ref{fig:oris11}, the full solution of the Bethe-Salpeter equation (\ref{bse}), using the simplified model from Ref. \cite{Inoue:2001ip}, is shown in red solid lines. The pole approximation from Eq. (\ref{pa}) is shown as the red dashed lines.
Indeed, the resonance shape is well described by the residue term, while there are higher order terms $a_0, \,a_1,\cdots$ in the expansion of the full amplitude, which provide an almost energy independent background. 

The $\pi N$ coupling is extracted at the pole position according to Eq. (\ref{pa}) and given in the isospin basis (baryon first coupling convention) by $g_{\pi N}(S_{11})=+\sqrt{2/3}|\pi^+n\rangle+\sqrt{1/3}|\pi^0p\rangle=0.68+0.39\,i$. This means that the residue, expressed as $a_{-1}=g_{\pi N}^2=|g_{\pi N}|^2\exp(2\,i\phi_{\pi N})$ has an angle of $2\,\phi_{\pi N}=60^0$ which is significantly different from zero and causes the distorted shape of the $N^*(1535)$ in Fig. \ref{fig:oris11} as compared to the classical resonance shape.

In particular, the real part of the $S_{11}$ amplitude (Re $S_{11}$) has a rise at energies around $z=1600$ MeV, which is rather the shoulder of the $N^*(1650)$ than part of the $N^*(1535)$. In other words, the original fit from Ref. \cite{Inoue:2001ip} tries to reproduce the narrow ``valley'' at Re $z=1550$~MeV, while this structure is tied to resonance interference~\cite{Doring2} which is not included in the model of Ref. \cite{Inoue:2001ip}. 
As a consequence, such a narrow structure requires a $N^*(1535)$ pole close to the physical axis; indeed, a width of $\Gamma=-2\,{\rm Im}\,z_0\sim 90$ MeV has been found in Ref. \cite{Inoue:2001ip} which is, although seen in some experiments~\cite{Bai:2001ua}, rather at the lower limit of realistic values for the width~\cite{Amsler:2008zz}.

Once we consider $\ezp$ in pion photoproduction instead of partial waves in $\pi N$ scattering, the phases change, as we will see in the following. In photoproduction, the residue $a_{-1}^\gamma$ is given by
\be
a_{-1}^\gamma&=&g_\gamma\,g_j,\non
g_\gamma&=&\sum_{i=1}^6 \tilde{\Gamma}^i g_i,\non
i\,\tilde{M}^{\rm PA}&=&\frac{a_{-1}^\gamma}{z-z_0}.
\label{paga}
\ee
with $\tilde{\Gamma}^i$ from Eq. (\ref{tfact}). The first line in Eq. (\ref{paga}) indicates that the residue can be decomposed in an effective photon coupling $g_\gamma$ and the strong coupling $g_j$ from Eq. (\ref{pa}) for the final meson-baryon state in channel $j$. In the absence of genuine resonance states, the phototransition coupling $g_\gamma$ is given by the sum over all photon loops, including the strong transition strength $g_i$, as the second line in Eq. (\ref{paga}) indicates. The sum is over the six channels in the particle basis as quoted in the first column of Table \ref{tab:su3c}. The pole approximation in photoproduction is then given by the third line in Eq. (\ref{paga}). 

The sum of the photon loops is over the coupled channels $\pi N$, $\eta N$, $K\Lambda$, and $K\Sigma$. This means the phototransition coupling $g_\gamma$ is sensitive to all the strong couplings $g_i$; they all appear in the transition, weighted by the respective photon loops. In particular, the residue $a_{-1}^\gamma$ will obtain a new phase that is, in principle, very different from the phase of the residue $a_{-1}$ of $\pi N$ scattering. 
Thus, the phase of a resonance in $\ezp$ can reveal valuable information about the transition strengths into the different coupled channels and their relative phases and magnitudes as they appear in the sum in Eq. (\ref{paga}). 

As a consequence of its dynamical generation, the\\ $N^*(1535)$ has large couplings to $K\Lambda$ and $K\Sigma$ \cite{Inoue:2001ip}, and it shows this feature in different theoretical approaches \cite{Kaiser:1995cy,Inoue:2001ip}. While the corresponding couplings cannot be accessed in direct experiments, they contribute to $g_\gamma$, and thus, to the phase of $\ezp$. Studying $\ezp$, therefore, provides a valuable tool to check the magnitudes and phases of the strong couplings to the dynamically generated $N^*(1535)$ pole and thereby helps confirm or rule out models of dynamical generation.  

In contrast to $\ezp$, observables like cross sections do not provide any information of the phase, or only indirectly through interference with other partial waves. Studying $\ezp$ provides, thus, a much more sensitive test in photonuclear reactions than in previous studies \cite{Jido:2007sm}.

We can determine the phase of the residue $a_{-1}^\gamma$ similar to the case of $\pi N$ scattering. In the case of $\ezp$, the phase is defined in the following way: we adopt a phase convention in which a phase of zero degrees corresponds to a classical resonance shape in $_p\ezp$ and $_n\ezp$. This means a single maximum in ${\rm Im}\,\ezp$ at the resonance position, while for ${\rm Re}\,\ezp$ a maximum below the resonance and a minimum above the resonance. Such a choice is fulfilled by the following definitions: 
\be
_p\tilde{a}_{-1}^\gamma&=&(-)\frac{-i\,M_N}{4\pi\,z}\,\left[\frac{\sqrt{2}}{3}\,a_{-1}^\gamma(n\pi^+)+\frac{1}{3}\,a_{-1}^\gamma(p\pi^0)\right],\non
_n\tilde{a}_{-1}^\gamma&=&(-)\frac{-i\,M_N}{4\pi\,z}\,\left[\frac{\sqrt{2}}{3}\,a_{-1}^\gamma(p\pi^-)-\frac{1}{3}\,a_{-1}^\gamma(n\pi^0)\right],\non
{\tilde \phi}^\gamma_{(p,n)}&=&\arctan\left({\rm Im}\,_{p,n}\tilde{a}_{-1}^\gamma/{\rm Re}\,_{p,n}\tilde{a}_{-1}^\gamma\right).
\label{phisezp}
\ee
These definitions of $_p\tilde{a}_{-1}^\gamma$ and $_n\tilde{a}_{-1}^\gamma$ take into account the connection between $\tilde M_{full}$ and $\ezp$ as given by Eqs. (\ref{ezp},\ref{isospinezp}). In Eq. (\ref{phisezp}) there is an additional minus sign that takes into account that a classical resonance shape is given by $-1/(z-z_0)$ and not $1/(z-z_0)$ [cf. Eq. (\ref{paga})]. 
In the calculation of the phases as defined in Eq. (\ref{phisezp}) an approximation is made by evaluating the $\tilde{d}$ functions appearing in $g_\gamma$ from Eq. (\ref{paga}) at $z={\rm Re}\,z_0$ instead at the pole position itself, $z=z_0$. 

The results for the phases of the $N^*(1535)$ in photoproduction, predicted by the model of Refs. \cite{Inoue:2001ip,Jido:2007sm}, are
\be
{\tilde \phi}^\gamma_{p}=178^0,\quad
{\tilde \phi}^\gamma_{n}=-9^0.
\label{angles}
\ee
The phase for $_n\ezp$ is $-9^0$; indeed, this value close to zero reflects the classical resonance shape as observed for $_n\ezp$ (red solid lines) in Fig. \ref{fig:oriprob}. For $_p\ezp$ the phase is $+178^0$, and we observe an inverted resonance shape (red solid lines) in Fig. \ref{fig:oriprob}. 

Thus, for the $N^*(1535)$ with $\gamma p$ or $\gamma n$ initial state, there is a relative phase of nearly $180^0$; this is in agreement with the findings of Ref. \cite{Jido:2007sm} where it was found that the helicity amplitudes $A_{1/2}^{(p)}$ and $A_{1/2}^{(n)}$ have opposite sign. In Ref. \cite{Jido:2007sm} this has been interpreted as a success of the model of Ref. \cite{Inoue:2001ip}, because this opposite sign (or nearly $180^0$ relative phase) is in agreement with the PDG values \cite{Amsler:2008zz}. Yet, we have seen here that there is more than a relative sign -- even with a $180^0$ relative phase, the predicted individual multipoles $_p\ezp$ and $_n\ezp$ strongly deviate from the partial wave analyses as Fig. \ref{fig:oriprob} shows.

Summarizing the findings of this section, the prediction for $\ezp$ using the original model of Ref. \cite{Inoue:2001ip} is seriously off the data. We have seen that there is already a potential problem in the description of the $\pi N\to\pi N$ scattering data: as the interference of the $N^*(1535)$ with the $N^*(1650)$ is neglected, the model from Ref. \cite{Inoue:2001ip} is forced to produce very small widths for the $N^*(1535)$. Second, for $\ezp$, the effective photon coupling $g_\gamma$ to the $N^*(1535)$, which is a model prediction, produces a phase of the\\ $N^*(1535)$ in photoproduction which is in disagreement with the partial wave analyses of the $\ezp$ multipole.

As discussed in Sec. \ref{sec:forma}, we will reconsider the model for the $N^*(1535)$ in Secs. \ref{sec:fit1}, \ref{sec:fit2} by allowing for the necessary degrees of freedom to solve the problems found in this section. This means the inclusion of the $N^*(1650)$ as a genuine resonance, and a second genuine resonance that is allowed to replace the dynamically generated one if this is required by the fit.

\subsubsection{Factorization of the phototransition}
\label{sec:factoam}
Discussed as case (IV) in Sec. \ref{sec:phomeba}, a further simplification of the phototransition amplitude can be carried out. We do not apply these further approximations in this study but discuss its consequences for the resonance phase. Case (IV) means an on-shell factorization of the photon loop. Then, in the absence of genuine resonances, the phototransition amplitude, as tested in this subsection, can be written as 
\be
\tilde{M}_{\rm fact.}^{\rm NP}=(1-V^\npo\, G)^{-1} v_\gamma^\mu\epsilon_\mu
\label{facall}
\ee  
where $v_\gamma^\mu$ is given by the set of tree-level diagrams $\gamma N\to MB$ with the meson-baryon pair $MB$ on-shell. The photon loop factorizes then into the form $v_\gamma^\mu\,G$ with $G$ being the factorized meson-baryon propagator. Eq. (\ref{facall}) implies that the regularization of this factorized $G$ is chosen to be the same as the $G$ in the hadronic part, for each channel $\pi N$, $\eta N$, $K\Lambda$, $K\Sigma$. Such a choice, together with the factorization, has been realized e.g. in Ref. \cite{Kaiser:1996js}. Additionally, an effective range expansion of the phototransition amplitude $v_\gamma^\mu$ has been carried out in Ref. \cite{Kaiser:1996js}. We test this simplified amplitude within the model of Ref. \cite{Inoue:2001ip}. Note that the hadronic interaction and the regularization (form factors) in the model of Ref. \cite{Kaiser:1996js} are different.

For the $\eta N$, $K\Lambda$, and $K\Sigma$ channels, the effective range expansion of $v_\gamma^\mu$ may be a sufficiently good approximation in the $N^*(1535)$ region due to the proximity of the respective thresholds. However, the $\pi N$ threshold is farther away from the $N^*(1535)$ region. 
While the effective range expansion is a good approximation up to 200 or 300 MeV above the $\pi N$ threshold, it is off by a factor of two in the $N^*(1535)$ region. We have seen in the previous section [cf. Eq. (\ref{paga})], that the resonance phase in photoproduction results from a subtle interference of the photon loops in the coupled channels; thus, this factor of two can cause large distortions in the $N^*(1535)$ phase.

Next, the full $E_{0+}$ amplitude is evaluated, using the effective range expansion for the on-shell factorized phototransition, and the hadronic amplitude from Ref. \cite{Inoue:2001ip}. With this treatment of the photon loops, the resonance angle is ${\tilde \phi}^\gamma_{p}=33^0$ which is more than $90^0$ different from the value of Eq. (\ref{angles}) of $178^0$.
In particular, the phase of $33^0$ is not too different from the phase in $\pi N\to\pi N$. However, as has been shown here, this should not be considered as a solution to the phase problem (cf. Fig. \ref{fig:oriprob}), but rather as a consequence of an oversimplified treatment of the photon loops. 

For the non-factorized photon loops used in this study, approximations have been made as well; however, as poin\-ted out at the end of Sec. \ref{sec:photoprod}, these approximations only change the phase by a few degrees, at least within the present framework [cf. end of Sec. \ref{sec:photoprod}]. As a further test, in Ref. \cite{Jido:2007sm} the photon loop has also been calculated relativistically. In this case, the phase of the phototransition amplitude changes by $17^0$ \cite{private_Jido} which is still small, compared to the phase problem found in the previous section.

\subsection{Pion production at low energies (Fit 1)}
\label{sec:fit1}
\begin{figure}
\includegraphics[width=0.483\textwidth]{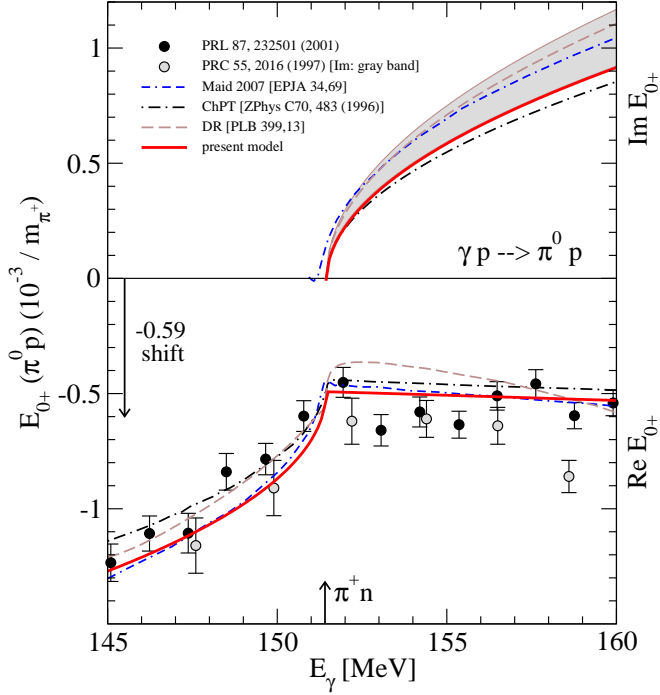}
\caption{(Fit 1, red lines); $\pi^0$ photoproduction close to threshold. The $\pi^+n$ threshold is indicated with an arrow. Experimental analyses: black (gray) data points from Refs. \cite{Schmidt:2001vg} (\cite{Bergstrom:1997jc}). Gray band from Ref. \cite{Bergstrom:1997jc}. Theory: Blue double dashed dotted lines: MAID 2007 \cite{Drechsel:2007if}, brown dashed lines: Ref. \cite{Hanstein:1996bd}. Dashed dotted lines: ChPT calculation from Ref. \cite{Bernard:1994gm} (see also the more recent work of Ref. \cite{Bernard:2005dj} and the analysis of Ref.~\cite{Kamalov:2001qg}).}
\label{fig10}
\end{figure}

In the following sections, we present the fit results using the model of the present work introduced previously (see Sec. \ref{sec:genureso} for the hadronic part and Sec. \ref{sec:formaphoto} for the phototransition amplitude).
Before considering the $N^*(1535)$ and $N^*(1650)$ region, it is instructive to study the low energy region. The low energy physics should not depend on resonances in the second resonance region, as required by chiral symmetry; this has been ensured by the use of derivative couplings for the bare $\pi N N^*$ and $\gamma N N^*$ vertices [cf. Eqs. (\ref{genterm}) and (\ref{nonre})]. Indeed, the corresponding parameters are very insensitive to the low energy region, as the fit shows. Thus, we have removed all contributions from the genuine resonances for the low energy fit. The only free parameters are then given by the four subtraction constants ($a_{\pi N}, a_{\eta N}, a_{K\Lambda}$ and $a_{K\Sigma}$) of the six coupled channels. Furthermore, those subtraction constants corresponding to the heavier channels, $K\Lambda$ and $K\Sigma$, are very insensitive to the low energy region, as expected. We have fitted the $S-$wave pion production, induced by photons and pions, up to $z\sim 1.4$ GeV  in energy. The results for the subtraction constants are shown in Table \ref{tab:parmsf1}; the resulting amplitudes are shown in Figs. \ref{fig10}, \ref{fig9}, and \ref{fig11}.
\begin{table}
\caption{Parameters of Fit 1 (low energy $\pi N\to\pi N$ and $\gamma N\to\pi N$). The parentheses indicate less influential parameters.}
\begin{center}
\begin{tabular}{rlrlrlrl}
 \hline\hline
$a_{K\Sigma}$&$(-3.80)$		&$a_{K\Lambda}$&$(3.80)$	&$a_{\pi N}$&$2.65$	&$a_{\eta N}$&$0.49$\\
\hline\hline
\end{tabular}
\end{center}
\label{tab:parmsf1}
\end{table}
We refer to this low energy fit as ``Fit 1'' in the following.

In Table \ref{tab:parmsf1}, the parentheses indicate the weak sensitivity of the subtraction constants of the heavy channels. The $\eta N$ subtraction constant only has an impact at the higher border of the considered energy interval. The value of $a_{\pi N}=2.65$ is comparable to that from the original model of Ref. \cite{Inoue:2001ip} of $a_{\pi N}=2.0$. 

The $\pi^0p$ photoproduction close to the $\pi N$ threshold offers a very sensitive test of the phototransition amplitude. Precise data exist for $\ezp (\pi^0 p)$, that include the cusp effect from different pion and nucleon masses. As the present model is formulated in the particle basis rather than the isospin basis, these effects can be taken into account in the present work. The result of Fit 1 for $\ezp (\pi^0 p)$ is shown in Fig. \ref{fig10} with the red solid lines. The experimental analyses are from Ref. \cite{Bergstrom:1997jc}. Note that for Im $\ezp$, only the experimental band from Ref. \cite{Bergstrom:1997jc} is shown. In the other analysis from Ref. \cite{Schmidt:2001vg}, a comparable value for Im $\ezp$ is obtained. In Fig. \ref{fig10}, also results from MAID are shown \cite{Drechsel:2007if,Hanstein:1996bd} and from a ChPT calculation \cite{Bernard:1994gm}.

The first thing to note is that $\ezp (\pi^0 p)$ is almost independent of the fit parameters. This is because the tree-level contribution and the one-loop contribution (i.e., the photon loop without rescattering) do not depend at all on the subtraction constants; the two-loop amplitude is orders of magnitudes smaller so close to threshold. 

Then, Im $\ezp (\pi^0 p)$ is entirely given by the $\pi^+ n$ one-loop amplitude. The strongly energy dependent cusp structure in Re $\ezp (\pi^0 p)$ is entirely given by the dispersive part of that loop. The tree-level photoproduction diagrams as well as the loop contribute to an almost energy independent background in Re $\ezp (\pi^0 p)$.

As Fig. \ref{fig10} shows, the photon loop evaluated in Sec. \ref{sec:photoprod} indeed predicts the correct energy dependence for both real and imaginary parts of $\ezp (\pi^0 p)$. This is a good test that our phototransition amplitude provides a realistic picture close to threshold.

We had to shift the theoretical result for Re $\ezp (\pi^0 p)$ by $-0.59\,\times 10^{-3}m_{\pi^+}^{-1}$, as indicated in the figure with an arrow. To judge the size of this amount, we consider the different contributions to Re $\ezp (\pi^0 p)$. From the tree-level diagrams of Fig. \ref{fig1}, only the direct and crossed nucleon exchange (b) and (d) contribute. For $S$-wave photoproduction, these diagrams are typically one order of magnitude smaller than the Kroll-Ruderman term (c), which vanishes identically for neutral pion photoproduction. This already shows the sensitivity of $\ezp (\pi^0 p)$ to higher order corrections. 

Still, these subleading contributions from Fig. \ref{fig1} (b), (d) are five times larger ($-2.47\,\times 10^{-3}m_{\pi^+}^{-1}$ ) than the experimental value at the $\pi^+ n$ threshold of $\sim$$ -0.5$$\times 10^{-3}m_{\pi^+}^{-1}$. The photon loop contributes with $+2.5$$\times 10^{-3}m_{\pi^+}^{-1}$; this almost fully cancels the tree-level contribution, so that the final theoretical result is around $0.59\,\times 10^{-3}m_{\pi^+}^{-1}$ larger than the experimental value. 

Thus, even the subleading tree-level diagrams (b), (d) vanish through cancellation by the photon loop. A part of the remaining discrepancy of $0.59\,\times 10^{-3}m_{\pi^+}^{-1}$
comes from the slightly inconsistent treatment of the tree level diagrams and the loop contributions, since the tree level is treated fully relativistically, while for the loop a non-relativistic framework is used.

\begin{figure}
\includegraphics[width=0.44\textwidth]{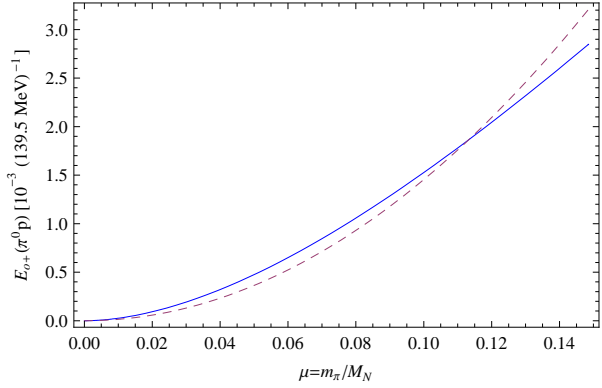}
\caption{Pion mass dependence of the triangle diagram. Solid line: Contribution to $E_{0+}(\pi^0 p)$ from the triangle diagram of Fig. \ref{fig:used_pholoops}(a) [meson pole term]. Dashed line: Nonanalytic piece of the NLO contribution given in Eq. (\ref{deltaeo}).}
\label{figchi}
\end{figure}

\begin{figure}
\includegraphics[width=0.483\textwidth]{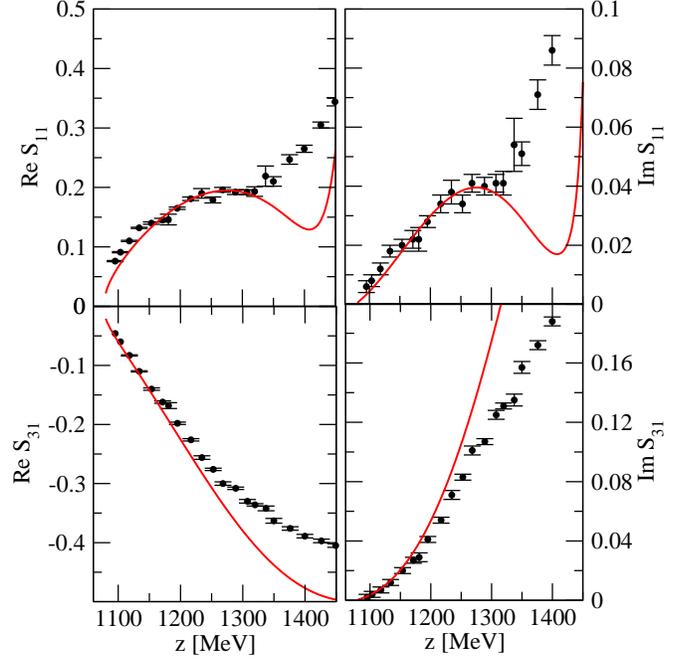}
\caption{(Fit 1) Low energy pion scattering. Analysis without resonances. Data from partial wave analyses as in Fig. \ref{fig:oris11}.}
\label{fig9}
\end{figure}

\begin{figure}
\includegraphics[width=0.483\textwidth]{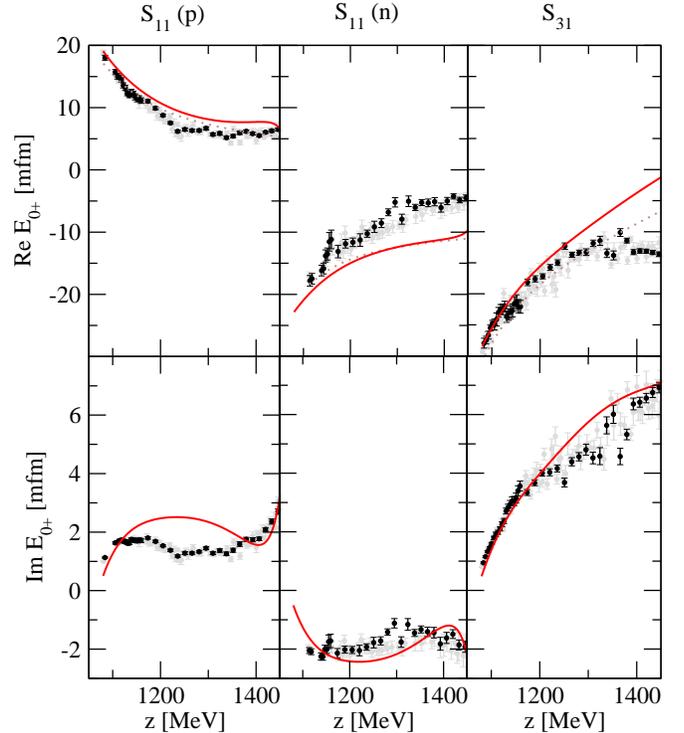}
\caption{(Fit 1) Low energy pion photoproduction. Analysis without resonances. Data from partial wave analyses as in Fig. \ref{fig:oriprob}.}
\label{fig11}
\end{figure}

Neutral pion production close to threshold has been calculated in the framework of chiral perturbation theory~\cite{Bernard:1994gm,Bernard:2005dj,Bernard:1991rt,Bernard:1992nc}. At next-to-leading (NLO) order, both the anomalous magnetic moment coming from the tree level diagrams and the triangle diagram contribute. The nonanlytical piece from the triangle diagram at NLO is given by~\cite{Bernard:1994gm,Bernard:2005dj}
\be
\Delta E_{0+}(\pi^0 p)=\frac{e g_A m_\pi^2}{128\,\pi^2 f_\pi^3}.
\label{deltaeo}
\ee
In Fig. \ref{figchi}, this pion mass dependence (dashed line) is compared to the contribution from the triangle diagram in the present formulation (solid line), given by Fig. \ref{fig:used_pholoops}(a). For this comparison, we have switched off the final state interaction $MB\to MB$, i.e., the shaded circle in Fig. \ref{fig:used_pholoops}(a) is given by the on-shell factorized Weinberg-Tomozawa term. Second, we consider only the $\pi N$ loop, and the tiny contribution from the photon coupling to intermediate $KY$ states is switched off.

The present result has the same $\mu\, (\equiv m_\pi/M_N)$ dependence as the result from Eq. (\ref{deltaeo}) (it does, though, include also higher order pieces) and both results seem to be in reasonable agreement as Fig. \ref{figchi} shows. Yet, comparing the result at the physical pion mass ($\mu=0.15$), what seems to be a small difference between both results, of $0.3\,\times 10^{-3}m_{\pi^+}^{-1}$, is of similar size as the previously stated discrepancy of $0.59\,\times 10^{-3}m_{\pi^+}^{-1}$, shown in Fig. \ref{fig10}. We can, thus, conclude that although the present calculation cannot precisely describe the data of Fig. \ref{fig10}, those discrepancies are much smaller than the NLO contribution in the chiral expansion ($\sim 3\,\times 10^{-3}m_{\pi^+}^{-1}$); neutral pion production close to threshold is difficult to describe as it is sensitive to higher-order corrections.

As final remark on $E_{0+}(\pi^0 p)$ near threshold, we noted that in the present work the real part of the photon loop is finite and fixed by gauge invariance, and there is no freedom from a regulator to adjust the model to the data.
In principle, the discrepancy in $E_{0+}$ discussed above may be eliminated by a proper adjustment to the data of the transverse contact current $T^\mu$ in Eq.~(\ref{eq:d9}), which has been set to zero in the present work. 
However, we have chosen not to do so in order to avoid introducing more free parameters in the current model.

The outcome of Fit 1 for $\pi N\to \pi N$ and $\gamma N\to \pi N$, fitted both in $S_{11}$ and $S_{31}$, is shown in Figs. \ref{fig9} and \ref{fig11}.
The results show a good global agreement with the ``data'' from the partial wave analyses, given that $a_{K\Lambda}$ and $a_{K\Sigma}$ are almost insensitive to this energy region; the fit thus describes, essentially with the two free parameters $a_{\pi N}$ and $a_{\eta N}$, ten different data sets. Still, there are remaining discrepancies in Re $S_{11}$ for $\pi N\to\pi N$ close to the $\pi N$ threshold, and for Im $\ezp$ on the proton. It is clear that the present model cannot deliver higher accuracy results at this point, because the interaction has been limited to the lowest order chiral interaction. It is known (see e.g. Ref. \cite{Doring:2004kt}) that close to the $\pi N$ threshold, higher order terms in the chiral interaction are needed to deliver a consistent threshold behavior within the present scheme of hadronic interaction. As this problem is well-known and our interest is different in the present study, we do not try to solve these discrepancies and refrain from calculating the scattering lengths.

In the next section, we will include higher energies in the fit. It is then not possible to fully maintain the quality of Fit 1 for the low energy region. We have already seen that one reason is the importance of higher orders in the chiral interaction. Second, the on-shell factorization scheme [cf. Sec. \ref{sec:hadpart}] leads to the appearance of subtraction constants, which are independent in energy. This energy independence is, of course, only an approximation, because higher order interactions induce more divergent loops, which require a multiple subtraction, i.e. a polynomial in energy instead of a constant. We thus have to recognize that the present scheme can only fit the data within an energy window, in which the subtraction polynomials are locally given by a constant value to a good approximation. Yet, while the model is limited at this point, one can still fit the entire $N^*(1535)$ and $N^*(1650)$ region to a sufficient precision while maintaining at the same time the main features of the low energy behavior of Fit 1.


\subsection{Pion production in the Second Resonance Region (Fit 2)}
\label{sec:fit2}
In this section, the main results of the present work are presented. After exploring the low energy region in the previous section, we now turn to the second resonance region employing the full model including the genuine resonances. The fitted data are the $S_{11}$ amplitude in $\pi N\to\pi N$ plus the $S_{31}$ amplitude below the region of the $\Delta(1620)$ (this resonance is not included in the present model). Simultaneously, the $S_{11}$ partial wave analysis data for $\ezp$ on the proton and on the neutron plus the $S_{31}$ $\ezp$ multipole below the $\Delta(1620)$ region is included. For all data on $\ezp$, we have only included the imaginary part of the amplitude; Re $\ezp$ is in all cases calculated but not included in the fit. This is because in the previous section we have seen that there is some theoretical uncertainty from higher order tree-level diagrams, which all contribute to the real part (up to unitarity corrections). In the fit, we have given more weight to the second resonance region, but also some weight to the low energy region to maintain the main features of the low energy Fit 1 discussed in the previous section.

The parameters of the solution are shown in Table \ref{tab:parmsf2}. There are four subtraction constants, and for each of the two genuine resonances, we have four bare hadronic couplings, two bare electromagnetic couplings, and one bare mass. We refer to this fit as ``Fit 2'' in the following.
\begin{table}
\caption{Parameters of Fit 2 ($\pi N\to\pi N$ and $\gamma N\to\pi N$, entire energy region). Subtraction constants, bare strong and electromagnetic couplings, and bare masses [cf. Eqs. (\ref{genterm}, \ref{nonre})]. The parentheses indicate less influential parameters.}
\begin{center}
\begin{tabular}{rlrlrl}
 \hline\hline
$a_{K\Sigma}$	&$-2.04$		&$g_8^{(1)}$			&\hspace*{0.25cm}$0.42$	&$g_8^{(2)}$			&\hspace*{0.25cm}$(2.73)$	\\
$a_{K\Lambda}$	&\hspace*{0.25cm}$3.80$	&$g_{8'}^{(1)}$			&$-0.03$		&$g_{8'}^{(2)}$			&\hspace*{0.25cm}$(1.21)$	\\
$a_{\pi N}$	&\hspace*{0.25cm}$1.29$	&$g_{\,\overline{10}}^{(1)}$	&$-0.21$		&$g_{\,\overline{10}}^{(2)}$	&\hspace*{0.25cm}$(0.42)$	\\
$a_{\eta N}$	&\hspace*{0.25cm}$0.93$	&$g_{27}^{(1)}$			&$-0.02$		&$g_{27}^{(2)}$			&$(-0.98)$			\\
		&			&$g_{\gamma pN^*}^{(1)}$	&\hspace*{0.25cm}$0.73$	&$g_{\gamma pN^*}^{(2)}$	&\hspace*{0.25cm}$(4.52)$	\\
		&			&$g_{\gamma nN^*}^{(1)}$	&$-0.44$		&$g_{\gamma nN^*}^{(2)}$	&$(-8.17)$			\\
		&			&$\bar{M}_{(1)}$\tiny{[MeV]}	&\hspace*{0.25cm}$1598$	&$\bar{M}_{(2)}$\tiny{[MeV]}	&\hspace*{0.25cm}$(3800)$	\\
\hline\hline
\end{tabular}
\end{center}
\label{tab:parmsf2}
\end{table}

\begin{figure}
\includegraphics[width=0.483\textwidth]{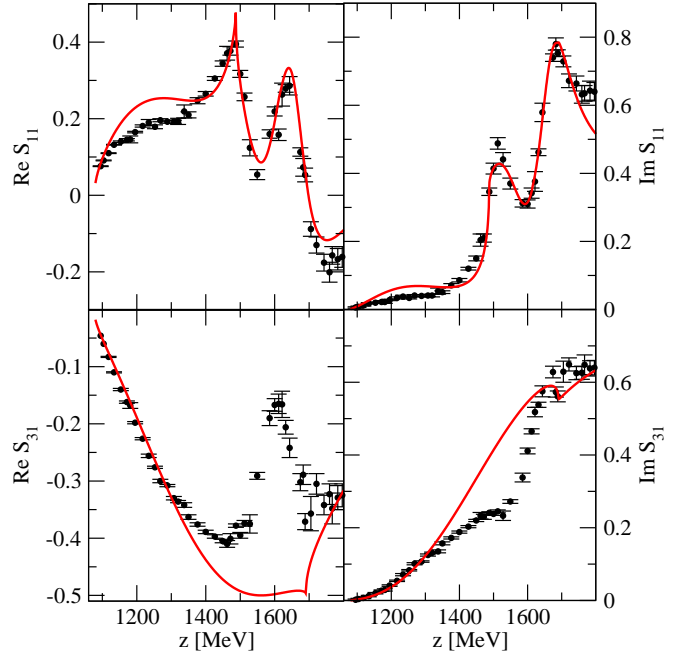}
\caption{Data points: SES $\pi N\to \pi N$ partial wave analysis from Ref. \cite{Arndt:2008zz} [FA07]. Red line: Joint analysis of $\pi N\to\pi N$ and $\gamma N\to\pi N$ (Fit 2).}
\label{fig2}
\end{figure}

\begin{figure}
\includegraphics[width=0.483\textwidth]{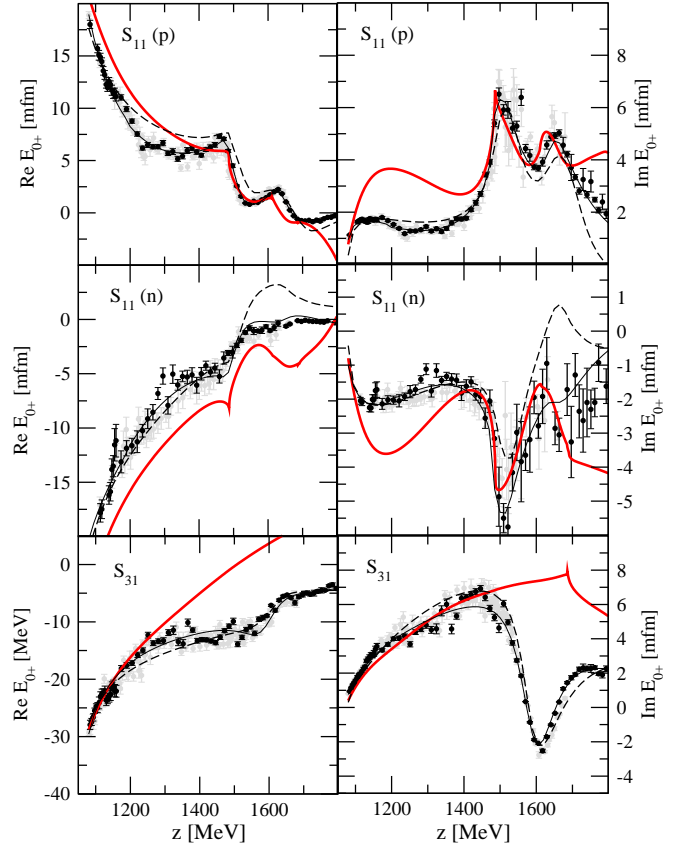}
\caption{Red lines: Joint analysis of $\pi N\to\pi N$ and $\gamma N\to\pi N$ (Fit 2). Data points and other lines: Labeling as in Fig. \ref{fig:oriprob}.}
\label{fig3}
\end{figure}

The resulting amplitudes are shown Figs. \ref{fig2} and \ref{fig3}. At low energies, the results are slightly worse than for Fit 1 (compare to Figs. \ref{fig9} and \ref{fig11}), while the threshold region for $\ezp(\pi^0p)$, shown in Fig. \ref{fig10} for Fit 1, is nearly unchanged. This is because the influence of resonances vanishes at the $\pi N$
threshold, as required by chiral symmetry. Second, the subtraction constants appear first at the two-loop level in photoproduction, which is very small close to threshold. 

As already discussed at the end of the previous section, one cannot maintain the quality of Fit 1, once the second resonance region is included in the fit. The reasons have been pointed out: higher orders in the chiral expansion are not included in the present scheme; closely connected to this, the subtraction polynomial is a constant in energy in the present model which restricts the freedom of the fit; third, the $\pi\pi N$ channel is not included. However, while higher order corrections will necessarily deliver a more precise fit, the qualitative results from the current model should not change. In this study, we are interested in the interplay of genuine and dynamically generated resonances, and the precision of the current model is sufficient for this discussion.

In any case, in the second resonance region, Fit 2 delivers a fair data description while maintaining the main features of Fit 1 at low energies. The results shown in Figs. \ref{fig2} and \ref{fig3} are much better than those from the original model of Ref. \cite{Inoue:2001ip}; in particular, the phase problem, pointed out in Sec. \ref{sec:phasprob} has been solved. Note also that the results are in fair agreement with Re $\ezp$, although only Im $\ezp$ has been included in the fit. 
The present results are in better agreement with data in the $N^*(1535)$ and $N^*(1650)$ region than in the previous work \cite{Inoue:2001ip} within the framework of U$\chi$PT.
In the $S_{31}$ partial-wave state, there are, of course, deviations in the region of the $\Delta(1620)$ as this resonance has not been included.

\begin{figure}
\includegraphics[width=0.43\textwidth]{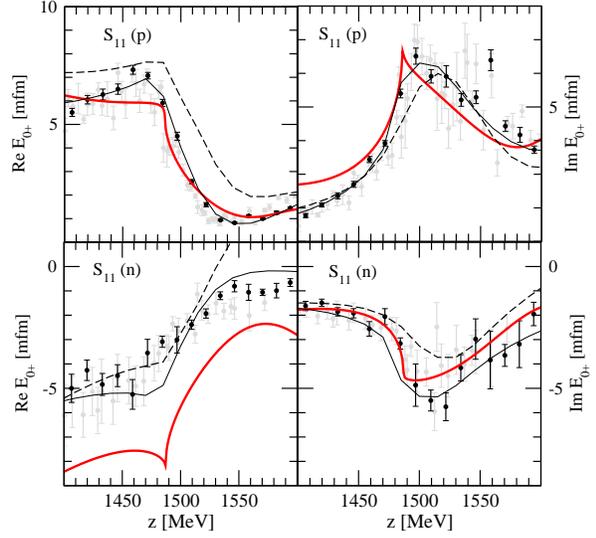}
\caption{Detail of Fig. \ref{fig3} in the $N^*(1535)$ region. Note the form of the cusp at the $\eta N$ threshold is quite different from the $\pi N\to\pi N$ reaction.}
\label{fig:zoom}
\end{figure}

Note the appearance of a cusp in Re $S_{11}$ in $\pi N\to\pi N$, while the imaginary part does not show such a pronounced cusp. In photoproduction, this is different: there are strong cusps both in the imaginary and real parts of $\ezp$ in $S_{11}(p)$ and $S_{11}(n)$ as Fig. \ref{fig3} shows. This different functional behavior is allowed as Watson's theorem does no longer hold above the $\pi\pi N$ threshold. Fit~2 reproduces these different functional forms. In fact, the previous SES analysis of $\ezp$ from Ref. \cite{Arndt:2002xv} (gray data in Fig. \ref{fig3}) did not clearly show the sharp cusp in Im $\ezp\,S_{11}(p)$, which first led to major concerns, because in the present model of dynamical generation of the $N^*(1535)$, that cusp is unavoidable and it is a very stable feature. However, the most recent analysis from Ref. \cite{Arndt:2008zz} (black data in Fig. \ref{fig3}) clearly reveals, that this cusp is indeed present, and very pronounced. This is shown in greater detail in Fig. \ref{fig:zoom}.

In Ref. \cite{Schutz:1998jx} the role of the $\eta N$ cusp in $\pi N\to\pi N$ has been discussed. The $N^*(1535)$ could be dynamically generated but at the cost of a worse data description and the appearance of a sharp pronounced cusp. It has then been argued that a genuine $N^*(1535)$ is preferred due to the better description of the $S_{11}$ partial wave. While in the present study, sharp cusps indeed appear, the data description is satisfactory; the dynamically generated $N^*(1535)$ is characterized by a true pole as will be seen in Sec. \ref{sec:discu}, and the cusp will be shown to be an interplay of physical and hidden poles, which necessarily appear. Note also, that in the latest version of the J\"ulich model \cite{Gasparyan:2003fp} a sharp $\eta N$ cusp appears despite of the fact that the $N^*(1535)$ is introduced as a genuine resonance state in that model. The worse data description in Ref. \cite{Schutz:1998jx}, for the case of the dynamical generation of the $N^*(1535)$, may be tied to the missing $K\Lambda$ and $K\Sigma$ channels in that model; including these channels in the J\"ulich model may help settle the issue of the nature of the $N^*(1535)$.

As argued in Sec. \ref{sec:photoprod}, we have used only the transverse part of the meson pole term for the phototransition loop amplitude. In contrast, the contribution from the baryon pole term [cf. Fig. \ref{fig:used_pholoops}] has been neglected. Indeed, we have checked that the contributions from this term are small. They are smaller than 10 \% for the photoproduction amplitude on the proton, and slightly larger than 10\% for the photoproduction amplitude on the neutron. In any case, one can safely neglect these contributions at the level of precision we are working in this study.

While both the $N^*(1535)$ and the $N^*(1650)$ are well reproduced by Fit 2, both in photon- and pion-induced reactions, we cannot conclude from the amplitude on the real $z$ axis, which role the genuine poles play. The values from Table \ref{tab:parmsf2} already indicate that the second genuine resonance has probably moved far into the complex plane (large bare couplings and mass). After all, only a study of the complex $z$ plane on unphysical sheets can resolve this issue. This will be carried out in the next section. 

Here, we anticipate some results of the next section. Indeed, in our final results (Fit 2), the $N^*(1535)$ is dynamically generated, while one of the genuine resonances is responsible for the $N^*(1650)$ and the other one provides a background that varies very slowly with energy. This ``background pole'' is far in the complex plane. As a test, we have refitted the data without the genuine resonance leading to this pole, but the $\chi^2$ becomes worse, indicating that the remaining free parameters of the model cannot compensate for the absence of the background pole. In order to see in which reaction this background is more important, we have performed further tests without second genuine resonance, fitting to (a) only $\pi N\to\pi N$ and (b) only $\gamma N\to\pi N$. In case (a), we can obtain a good fit, with a dynamically generated $N^*(1535)$ and a genuine $N^*(1650)$, without the need for an additional background. In case (b), however, we cannot obtain a good fit; in the best fit, the only remaining genuine state even becomes a background pole (i.e. moves far into the complex plane), instead of describing the $N^*(1650)$. 

Thus, we can conclude that, particularly in photoproduction, an additional background is required on top of the $N^*(1535)$ and $N^*(1650)$ and the background generated by the rescattering provided by Eq. (\ref{bse}) from $V^\npo$. Such an additional background in photoproduction could be provided by diagrams which are not explicitly included in the present study, like vector meson $t$-channel exchange with anomalous photon couplings.

Finally, we have also studied the importance of the $K\Lambda$ and $K\Sigma$ channels for the formation of the $N^*(1535)$. It is known that these channels are responsible for the necessary attraction to generate the $N^*(1535)$ \cite{Kaiser:1995cy,Kaiser:1996js,Inoue:2001ip}. In a simple model without genuine resonances, we have fitted the $\pi N\to\pi N$ reaction, only in $S_{11}$ and only around the resonance position. Additionally, we have increased $f_K$ in Eq. (\ref{wtt}) from its original value of $f_K=1.22\,f_\pi=113.5$ MeV (a more precise value of $f_K=1.193\,f_\pi$ has been reported recently in Ref. \cite{Bernard:2007tk}); this models a weaker coupling to the strangeness channels $K\Lambda$ and $K\Sigma$. Within a reasonable range for the subtraction constants (maximum 3 to 4), we could get a resonant shape of the $N^*(1535)$ for values of $f_K$ up to 150 or 160 MeV. For larger $f_K$, the resonance fades away. A maximum value of $f_K=150$ MeV has been regarded in Ref. \cite{Borasoy:2007ku} as a reasonable limit.

Thus, we can conclude that the dynamical generation of the $N^*(1535)$ indeed requires a sufficiently strong coupling to the $K\Lambda$ and $K\Sigma$ channels. Since in the present approach, the coupling strengths to these channels are given by the SU(3) Lagrangian which are strong, the dynamically generated $N^*(1535)$ appears as quite a stable structure
in the various refits discussed in this work.

More data can be included in the fit that allow to further test the SU(3) structure, such as those from the pion- and photon-induced $\eta N$, $K\Lambda$, and $K\Sigma$ production, as well as from the corresponding electroproduction processes. As we have 18 free parameters altogether, an inclusion of those data is appropriate to impose further constraints on the model. The results for these observables will be presented in Ref. \cite{inprep}; the current Fit 2 already delivers a good qualitative agreement for these observables. 


\section{Discussion}
\label{sec:discu}

\subsection{Pole positions and residues}
\label{sec:analcont}
Poles and zeros of the amplitude in the complex plane of the scattering energy $z\equiv s^{1/2}$ determine the global appearance of the amplitude on the physical axis.
The hadronic amplitude can be analytically continued to the complex $z$-plane. There are two different Riemann sheets for each channel $\pi N$, $\eta N$, $K\Lambda$, and $K\Sigma$~\cite{Doring2}. 

The first sheet of the propagator $G$ is defined by evaluating $G$ for complex $z$, while the second sheet is obtained by adding twice the discontinuity of $G$ along the right-hand cut,
\be
G^{(1)}(z)&=&G(z),\non
G^{(2)}(z)&=&G(z)+2\,\frac{i\,M\,q_{\rm{on}}^>}{4\pi\,z}\ ,
\label{sheetg}
\ee
where $M$ is the baryon mass of a given channel and 
\be
q_{\rm{on}}^>&=&
\begin{cases}
-q_{\rm{on}}	&	\text{if Im $q_{\rm{on}}<0$}\\
\,\,q_{\rm{on}}	&	\text{else}
\end{cases}
\label{sq_pres}
\ee
is the on-shell relative momentum. Eq. (\ref{sq_pres}) ensures that $q_{\rm{on}}^>$ has the cut along the right-hand side. With this prescription, both $G^{(1)}$ and $G^{(2)}$ have the cut along the positive physical axis and are analytically connected with each other along these cuts.

The various sheets of the scattering amplitude $T$ are induced by the replacement of $G$ from Eq. (\ref{bse}) with $G^{(1)}$ or $G^{(2)}$. This amounts, for the four meson-baryon channels, to 16 sheets of the scattering amplitude (in this section, we work in the isospin limit, so that the 6 channels in the particle basis reduce to the four channels $\pi N$, $\eta N$, $K\Lambda$, and $K\Sigma$). Out of these 16 sheets, only a few are directly connected to the physical axis. Directly connected in this sense means connected without having to turn around branch points to reach the physical axis. This issue is discussed in detail in Ref. \cite{Doring2}. 

In the following, the physical sheet of $T$ induced by $G^{(1)}$ is labeled 1, the second, or unphysical sheet, induced by $G^{(2)}$, is labeled 2, for a given channel. For example, in the channel ordering $\pi N$, $\eta N$, $K\Lambda$, $K\Sigma$, sheet 1111 is the physical sheet with respect to all channels. It is free of poles. Sheet 2211 is the sheet that is given by the unphysical sheet of $\pi N, \,\eta N$, and the first sheet of $K\Lambda$ and $K\Sigma$. In the following, we concentrate on the lower $z$ half plane. The properties of the amplitude in the upper $z$ half plane are analogous to those of the lower half plane and $T^*(z)=T(z^*)$ (Schwartz's reflection principle).

\begin{figure}
\includegraphics[width=0.4\textwidth, height=0.9\textheight]{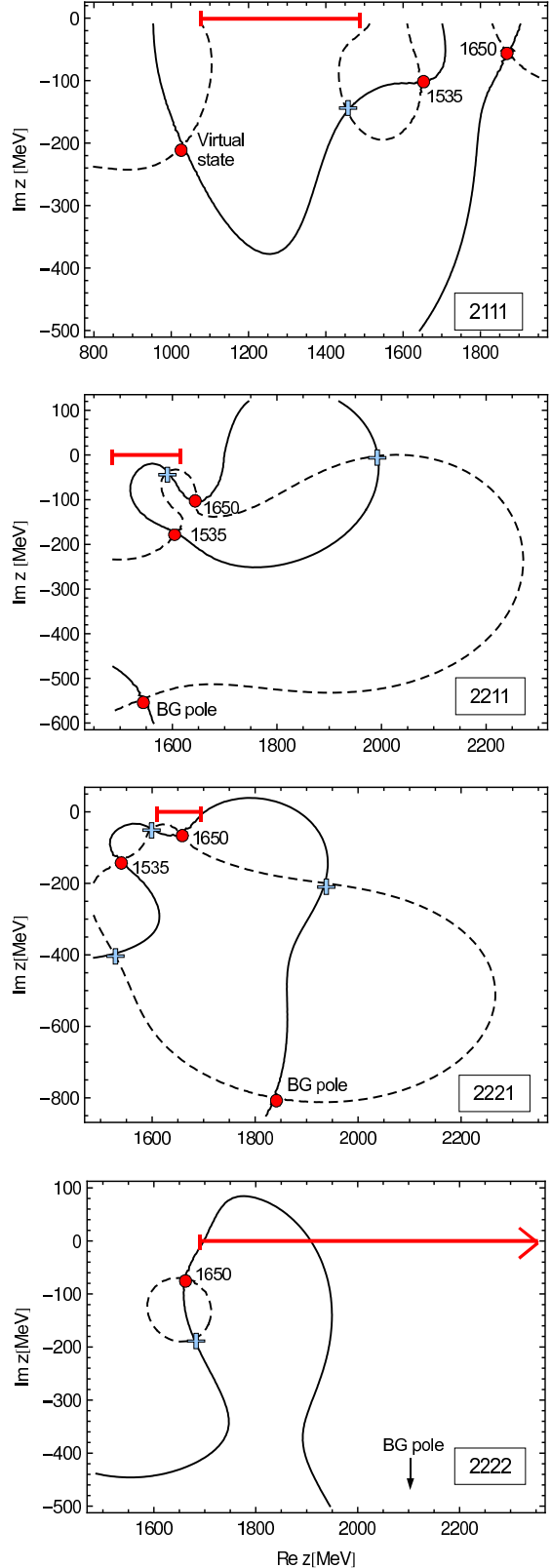}
\caption{``Gau{\ss} plot'' of the Riemann sheets 2111 to 2222. The contours Re $T=0$ (solid lines) and Im $T=0$ (dashed lines) intersect at poles (red circles) and zeros (blue crosses) of the amplitude. The part of the physical axis directly connected to the respective sheet is indicated in bold red.}
\label{fig:group_gauss}
\end{figure}

As discussed in Ref. \cite{Doring2}, for a given channel, the first sheet in the lower $z$ half plane is directly connected to the physical axis below the threshold $z<m_i+M_i$. In contrast, the physical axis above threshold $z>m_i+M_i$ is directly connected to the second sheet in the lower $z$ half plane. For example, the sheet 2111 is directly connected to the physical axis for $m_\pi+M_N<z<m_\eta+M_N$. Thus, there are four combinations of sheets that are directly connected to the physical axis, 2111, 2211, 2221, 2222. The corresponding pieces of the physical axis for these sheets are indicated in Fig. \ref{fig:group_gauss} with the bold red lines.

Usually, poles are only searched on one combination of the sheets
\cite{Inoue:2001ip,Sarkar:2004jh} using the prescription for the propagator $\bar{G}_i$
\be
\bar{G}_i(z)=\begin{cases}
G_i^{(1)}(z)	&	\text{if Re $z<m_i+M_i$}\\
G_i^{(2)}(z)	&	\text{if Re $z\geq m_i+M_i$}.
\end{cases}
\label{choi2}
\ee
for the channels $i$ with meson and baryon mass $m_i$ and $M_i$, respectively. 
The prescription given by Eq. (\ref{choi2}) is a sensible one, because the Riemann sheet induced by $\bar{G}(z)$ is the closest one to the physical axis. Poles on other sheets are expected to have much less impact on the amplitude on the physical axis.

However, studying all four combinations 2111 to 2222 instead of using $\bar{G}$ from Eq. (\ref{choi2}) reveals also virtual states and hidden poles, that are not found using the prescription from Eq. (\ref{choi2}). Those sub- and above-threshold resonances can have a large influence on the physical amplitude as we will see in the following.

In Fig. \ref{fig:group_gauss}, we show contour plots of the amplitude for the contours Re $T(z)=0$ (solid lines) and Im $T(z)=0$ (dashed lines). It is easy to see that those lines intersect at poles (red circles) and zeros (blue crosses) of the amplitude. In the following, we refer to this kind of representation of a complex function as ``Gau{\ss} plot''. 

The pole positions and the couplings to the $\pi^-p$ channel, defined in Eq. (\ref{pa}), are listed in Table \ref{tab:poles}. 
\begin{table}
\caption{Positions and couplings $g_{\pi^-p}$ [cf. Eq. (\ref{pa})] of poles on different sheets. VS means the virtual state below the $\pi N$ threshold, BG means the background pole. The physical poles are underlined [cf. discussion in Sec. \ref{sec:discusheets}]. The virtual state and the $N^*(1535)$ appear dynamically generated, while the $N^*(1650)$ and the background pole result from genuine resonance states.}
\begin{center}
\begin{tabular}{lll}
 \hline\hline
		&Position [MeV]\,\,\,\,\,&$g_{\pi^-p}$		\\
Sheet 2111\,\,\,\,\,\,&		&			\\
\,\,\,VS	&$1031-203\,i$	&$-0.51+1.58\,i$	\\
\,\,\,$N^*(1535)$	&$1647-103\,i$	&$-1.55+1.40\,i$	\\
\,\,\,$N^*(1650)$	&$1872-57\,i$	&\,\,\,\,\,$0.91+2.64\,i$\\
Sheet 2211	&&\\
\,\,\,\underline{$N^*(1535)$}	&$1608-175\,i$	&\,\,\,\,\,$3.35+1.82\,i$\\
\,\,\,$N^*(1650)$	&$1645-105\,i$	&$-1.83+1.88\,i$	\\
\,\,\,BG	&$1545-545\,i$	&$-0.78+3.52\,i$	\\
Sheet 2221	&&\\
\,\,\,$N^*(1535)$	&$1538-139\,i$	&\,\,\,\,\,$1.42+0.46\,i$\\
\,\,\,\underline{$N^*(1650)$}&$1655-59\,i$	&$-0.89+0.48\,i$	\\
\,\,\,BG	&$1837-800\,i$	&\,\,\,\,\,$0.31+2.39\,i$\\
Sheet 2222	&&\\
\,\,\,$N^*(1535)$	&no pole	&			\\
\,\,\,$N^*(1650)$	&$1662-72\,i$	&$-1.03+0.12\,i$	\\
\,\,\,BG	&$2129-1289\,i$	&\,\,\,\,\,$0.33+2.26\,i$\\
\hline\hline
\end{tabular}
\end{center}
\label{tab:poles}
\end{table}
The coupling strengths to isospin $I=1/2$ can be obtained through $g_{I=1/2}=\sqrt{3/2}\,g_{\pi^-p}$ up to small isospin breaking from different masses which is only important for the cusp effect in $\ezp(\pi^0p)$, shown in Fig. \ref{fig10}.
The couplings to the other channels $g_{\eta N}$, $g_{K\Lambda}$, and $g_{K\Sigma}$ are not listed in Table \ref{tab:poles}, although they have been calculated. They will be fully quoted in Ref. \cite{inprep}, once the observables in the $\eta N$, $K\Lambda$, and $K\Sigma$ channels are included in the fit.

There are two genuine poles in the model. One has become the $N^*(1650)$ in the solution, the other one has become a background pole that lies far in the complex plane and provides an almost energy independent background. The virtual state VS (discussed below) and the $N^*(1535)$ appear as dynamically generated. 

The couplings of the $N^*(1535)$ on sheet 2211 to $K\Lambda$ and $K\Sigma$ are $|g_{K^0\Lambda}|=4.3$ and $|g_{K^+\Sigma^-}|=2.3$. The large coupling to $K\Sigma$ indicates that the $N^*(1535)$ appears as a quasibound $K\Sigma$ state, in qualitative agreement with the models of dynamical generation from Refs. \cite{Kaiser:1995cy,Kaiser:1996js} and \cite{Inoue:2001ip}. Yet, the situation is more complicated here, because the interference with the $N^*(1650)$ is nontrivial, as will be discussed in Sec. \ref{sec:discusheets}. Second, the coupling strengths $g$ of the $N^*(1535)$ on sheet 2211 --which is the pole to be compared with the pole of Ref. \cite{Inoue:2001ip}-- are larger than those from Ref. \cite{Inoue:2001ip}, for all channels. However, this is a simple consequence of the fact that the $N^*(1535)$ is much wider here than in Ref. \cite{Inoue:2001ip}. In order to achieve a comparable resonance shape on the physical axis at Im $z=0$, a pole located farther in the complex plane needs to have a large residue according to Eq. (\ref{pa}).

\subsection{Discussion of four sheets}
\label{sec:discusheets}

In Fig. \ref{fig:group_gauss}, consider first sheet 2211, which is connected to the physical axis in the range $m_\eta+M_N<z<m_K+M_\Lambda$. 
This sheet shows the global behavior of the solution: the $N^*(1535)$ is dynamically generated with a large width, the first genuine pole is identified with the $N^*(1650)$, and the second genuine pole has moved far into the complex plane, providing a background that varies very slowly with energy on the physical axis (``background pole''). The latter pole models additional background processes that are not explicitly included in the present model. 
Thus, there are several poles and zeros on sheet 2211. One of the zeros is situated in between the $N^*(1535)$ and $N^*(1650)$. This zero is also found in Refs. \cite{Arndt:2003if} and \cite{Doring2}. The other zero is near $z=2000-0\,i$ MeV. As it lies above the $K\Lambda$ and $K\Sigma$ thresholds, it is physically not observable; a zero on that position could only be observable on sheet 2222. See also a discussion in Ref. \cite{Gasparyan:2003fp} on a zero on the physical axis in the $\pi N\to\eta N$ transition in the $S_{11}$ state.  

What are the physical implications of the poles on sheet 2211? Sheet 2211 is connected to the physical axis in the range $m_\eta+M_N<z<m_K+M_\Lambda$. The pole of the $N^*(1535)$ is within this energy window. We can identify this pole with the physically observable $N^*(1535)$. The $N^*(1535)$ pole lies relatively far in the complex plane at Im $z=-175$ MeV; in the recent work of Ref. \cite{Suzuki:2008rp}, a similarly large imaginary part for the $N^*(1535)$ pole has been found (Im $z_0=-191$ MeV) while the $N^*(1535)$ pole found in Ref. \cite{Doring2} is much closer to the physical axis (Im $z_0=-64.5$ MeV).

The real part of the pole position of the $N^*(1650)$ on sheet 2211 is above $m_K+M_\Lambda$. Thus, the $N^*(1650)$ on sheet 2211 is rather an above-threshold resonance. Only its low-energy tail is visible on the physical axis. The physical $N^*(1650)$ has to be searched for on sheet 2221 instead (see below).

Sheet 2111 in Fig. \ref{fig:group_gauss} is connected to the physical axis in $m_\pi+M_N<z<m_\eta+M_N$. Below the $\pi N$ threshold and far from the real axis into the complex plane, there is a pole on sheet 2111 as Fig. \ref{fig:group_gauss} shows. We have found this virtual state in the $S_{11}$ partial wave also in other models \cite{Doring2}; as discussed below, it is connected to the sharp rise of Re $S_{11}$ at the $\pi N$ threshold and seems to be required by the partial wave. However, it is not clear if this state is genuine or a ``forced'' pole that mocks up the $u$- and $t$-channel subthreshold cuts that are not explicitly included in the present model~\cite{privcom}.

The sheets 2221 and 2222 are also shown in Fig. \ref{fig:group_gauss}. They are connected to the physical axis within the ranges $m_K+M_\Lambda<z<m_K+M_\Sigma$ and $m_K+M_\Sigma<z<\infty$, respectively. The structure of sheet 2221 is similar to that of 2211. The pole of the $N^*(1650)$ on sheet 2221 is directly connected to the physical axis, for which we can identify it with the physical $N^*(1650)$. The physical $N^*(1535)$ and $N^*(1650)$ on their respective sheets are highlighted in Table \ref{tab:poles}. As discussed before, the other $N^*$ poles on other sheets always appear either as a sub-threshold or above-threshold resonance. Yet, also these secondary poles are important, because their tails can be visible on the physical axis, and their interference with the physical poles is important.

 The structure of sheet 2222 is quite different from those of 2211 and 2221. Sheet 2222 is the sheet connected to the physical axis above the $K\Sigma$ threshold. The $N^*(1535)$ has disappeared on that sheet and the background pole has moved even farther into the complex plane (out of the plotted range, see Table \ref{tab:poles}).

The disappearance of the $N^*(1535)$ pole on sheet 2222 is an interesting fact; the model of dynamical generation of the $N^*(1535)$ from Ref. \cite{Inoue:2001ip} shows the same behavior. The absence of the $N^*(1535)$ at high energies has implications for the concept of sub-threshold resonances that are sometimes used to fit reaction data: for example, in Ref. \cite{Xie:2007qt}, the role of the $N^*(1535)$ in the reaction $\pi^-p\to\phi N$ was discussed. There, the $N^*(1535)$ was used as a sub-threshold resonance to explain the $\phi N$ cross section, i.e., the resonance was extrapolated several hundreds of MeV above its position. It was already argued in Ref. \cite{Doring:2008sv}, that one should rather use the full energy dependent meson-baryon amplitude instead of extrapolating the resonance; with the detailed study of the Riemann sheets done here, we can further sharpen this statement: the influence of the $N^*(1535)$ has just disappeared completely above the $K\Sigma$ threshold. 

The disappearance of the $N^*(1535)$ above the $K\Sigma$ threshold is, of course, tied to the present model. Yet, even in cases the $N^*(1535)$ does not disappear from the 2222 sheet, due to its strong coupling to the strangeness channels (implicitly assumed when used to fit the $\phi N$ production), its position will unavoidably change drastically on the 2222 sheet compared to its position on the 2211 sheet, where it is observed. This is a model-independent behavior, and makes any kind of models using phenomenological sub-threshold resonances questionable. 

\subsection{The impact of poles on the physical axis}

\begin{figure}
\includegraphics[width=0.4\textwidth]{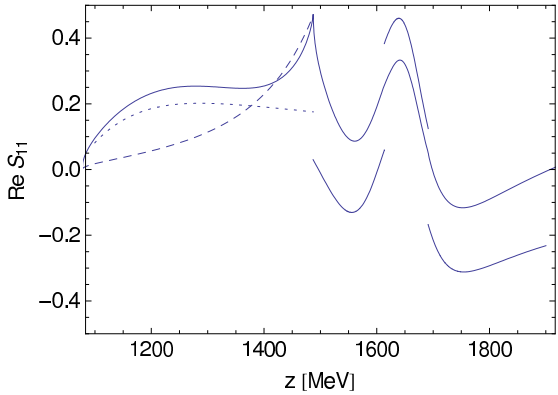}
\caption{Pole approximations $T_{\rm PA}^{(i)}$ (piecewise defined lines) of the full solution for Re $S_{11}$. The pieces of the PA are limited by the thresholds of the channels $\pi N$, $\eta N$, $K\Lambda$ and $K\Sigma$.}
\label{fig:pa}
\end{figure}

The various poles on different sheets have different impact at the physical axis. Consider the pole approximation from Eq. (\ref{pa}), that provides the leading term in the Laurent expansion around the pole position. In Fig. \ref{fig:pa} we shows the real part of the $S_{11}$ amplitude together with the expansions from Eq. (\ref{pa}).

Below the $\eta N$ threshold ($z<1487$ MeV), two curves are shown in Fig. \ref{fig:pa}. The dotted curve originates from the virtual state below the $\pi N$ threshold on sheet 2111 shown in Fig. \ref{fig:group_gauss} [cf. Table \ref{tab:poles}]. It almost saturates the amplitude close to the $\pi N$ threshold, which indeed shows that the low energy region is dominated by this virtual state.  

At the $\eta N$ threshold, the Re $S_{11}$ amplitude shows a characteristic cusp. The shape above the  $\eta N$ threshold is well approximated by the poles on the sheet 2211, while below the $\eta N$ threshold, it is well described by the ``hidden'' $N^*(1535)$ pole on the 2111 sheet, as indicated in Fig. \ref{fig:pa} with the dashed line. Before studying the other pole approximations above the $\eta N$ threshold, plotted in Fig. \ref{fig:pa}, we further discuss the cusp.

\begin{figure}
\includegraphics[width=0.47\textwidth]{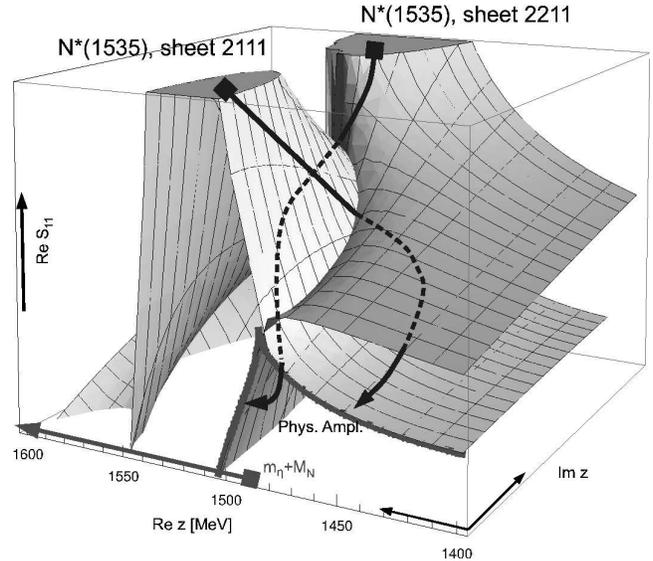}
\caption{Cusp of the physical amplitude (schematically). Above the $\eta N$ threshold, the amplitude is dominated by the physical $N^*(1535)$ pole on sheet 2211. Below the $\eta N$ threshold, the amplitude is dominated by the hidden $N^*(1535)$ pole on sheet 2111.}
\label{fig:cusp3d}
\end{figure}
In Fig. \ref{fig:cusp3d} the cusp and its origin are schematically displayed. The physical amplitude at Im $z=0$, around the cusp, is connected to two different Riemann sheets, 2111 and 2211. The $N^*(1535)$ poles are at different positions on these sheets, and their influence on the amplitude, indicated with thick arrows, results in a cusp. This explains naturally the appearance of the cusp.

The physical axis above the $\eta N$ threshold is divided into three pieces, separated by the $K\Lambda$ and $K\Sigma$ thresholds. The pieces are connected to sheets 2211, 2221, and 2222, respectively. We have summed the contributions from the poles from Fig. \ref{fig:group_gauss} for each of these sheets according to
\be
T_{\rm PA}^{(i)}=\sum_j \frac{a_{-1}^j}{z-z_0^j}
\ee
where $(i)$ indicates the amplitude on sheet $i=$2211, 2221, 2222 and the sum is over the poles on a given sheet. The different $\pi N\to\pi N$ residues and pole positions on a given sheet $(i)$ are indicated as $a_{-1}^j$ and $z_0^j$. 
It is important to sum over the different poles. In particular, the combination of the $N^*(1535)$ and $N^*(1650)$ poles is important. This is because the $N^*(1535)$ provides a strongly energy-dependent background for the $N^*(1650)$ and vice versa. In other words, the interference between these two resonances is responsible for the shape of the partial wave amplitude; they cannot be treated separately. This issue has been extensively discussed in Ref. \cite{Doring2}.
 
The resulting pole approximations $T_{\rm PA}^{(i)}$ are shown in Fig. \ref{fig:pa} with the solid lines. They are different for each region of the physical axis between the channel thresholds. The results provide a good reproduction of the energy dependence of the amplitude, in particular the resonance shapes of the $N^*(1535)$ and $N^*(1650)$. Yet, for each piece of the physical axis above the $\eta N$ threshold, the pole approximations are still off the full solution by backgrounds that slowly vary with energy. These almost constant off-sets, that are different for each piece, come from higher order terms in the Laurent expansion around the pole positions, $a_0$, $a_{1}$ according to Eq. (\ref{pa}).

Finally, let us mention our results of the pole search in the model of Ref. \cite{Inoue:2001ip}. In that reference, only the dynamically generated pole of the $N^*(1535)$ on sheet 2211 has been found. However, we can perform the same detailed pole search as carried out for the present model. Then, one indeed finds more poles in the model of Ref. \cite{Inoue:2001ip}. As in the present study, there is another pole of the $N^*(1535)$ on sheet 2111. Similarly to the present model, the cusp structure at the $\eta N$ threshold can be well described by these two poles. Also, the virtual state below the $\pi N$ threshold is present in the model of Ref. \cite{Inoue:2001ip}. In the same way as found here, that state is responsible for the sharp rise of the real part of the $S_{11}$ amplitude close to threshold. Furthermore, the model of \cite{Inoue:2001ip} has another dynamically generated resonance on the sheet 2111 far in the complex plane at $z_0=1657-267\,i$ MeV that is responsible for some of the structure along the physical axis between the $\pi N$ and $\eta N$ thresholds.

\subsection{The phase problem revisited}
In light of the analysis of Fit 2, the phase problem in the model of Ref. \cite{Inoue:2001ip}, found in the present study and discussed in Sec. \ref{sec:phasprob}, can be revisited. As we have seen in Figs. \ref{fig2}, \ref{fig3}, and \ref{fig:zoom}, the amplitude of Fit 2 shows no phase problem any more on the physical axis. In the previous section, we have seen that the physical axis is dominated by the different $N^*(1535)$ and $N^*(1650)$ poles on the different sheets. In particular, one has to consider the interference of resonances, and individual contributions make no physical sense. In view of this, one should be cautious to quote phase angles for the $N^*(1535)$ alone.

Yet, the interfering poles dominate the energy dependence in $\gamma N\to\pi N$ and $\pi N\to\pi N$, as has been discussed. In particular, the phases of the resonances, together with the residue strengths and pole positions, lead to very different resonance shapes as a comparison of Figs. \ref{fig2} and \ref{fig3} shows. Most noticeably, the different strengths of the $\eta N$ cusps in the real and imaginary parts of the $\gamma N\to\pi N$ and $\pi N\to\pi N$ amplitudes can be explained by Fit 2: the interfering resonances with different phases modify the amplitudes in such a way, that the different functional forms turn out naturally.

In particular, the $N^*(1535)$ shape appears narrower in $\ezp$ than in $\pi N\to\pi N$ (see e.g. the narrow width found in Ref. \cite{Drechsel:2007if}). However, in the present framework, this could be explained naturally by the photon coupling to the dynamically generated $N^*(1535)$ which induces a different phase [cf. Eq. (\ref{paga})] on the coupling constant. With such a new phase in photoproduction, together with the resonance interference, the $N^*(1535)$ can naturally appear narrower on the physical axis, while its pole position is, of course, still the same to that in $\pi N\to\pi N$.


\section{Conclusions}
The $N^*(1535)$ has been previously described as a purely dynamically generated resonance from the unitarized lowest order chiral interaction in $SU(3)$ coupled channel dynamics. This concept has been tested in a variety of pion- and photon-induced reactions. In this study, we carry out a further test, which is more sensitive  because it is directly tied to the amplitudes instead to cross sections. This is a very sensitive test to get further insight into the nature of resonances. 

The simultaneous study of the reactions $\pi N\to\pi N$ and $\gamma N\to\pi N$ for the $S_{11}$ and $S_{31}$ partial waves reveals a phase inconsistency of the $N^*(1535)$ in the previous description of dynamical generation of Ref. \cite{Inoue:2001ip}. Part of the phase inconsistency could be traced back to the absence of the $N^*(1650)$ resonance which strongly affects the properties of the $N^*(1535)$ through resonance interference. Thus, in an extension of the original model of Ref. \cite{Inoue:2001ip}, we allow for two genuine 3-quark resonances: one to account for the $N^*(1650)$, and another one to replace the dynamically generated $N^*(1535)$, if the fit prefers this solution. 

In this work, we first show a fair agreement of the present model with the data close to threshold for $\pi^0p$ photoproduction. In the fit of the $N^*$ region, one of the genuine poles indeed accounts for the $N^*(1650)$. However, the other one, instead of replacing the dynamically generated $N^*(1535)$, moves far into the complex plane and provides an almost energy independent background; it accounts for background processes not explicitly included in the present photoproduction model such as $t$-channel vector meson exchanges with anomalous photon couplings. 

In any case, the dynamically generated $N^*(1535)$ pole appears as a stable object, resistant to changes of the fit. While its position changes significantly compared to the original model of Ref. \cite{Inoue:2001ip}, due to an interference with the $N^*(1650)$, the present study shows that a dynamically generated $N^*(1535)$, together with a genuine $N^*(1650)$, can deliver a consistent picture simultaneously in $\pi N\to\pi N$ and $\gamma N\to\pi N$. 

The dynamical generation is tied to the strong couplings to the $K\Lambda$ and $K\Sigma$ channels provided by the SU(3) Lagrangian. In this connection, we have also verified that the $N^*(1535)$ as a dynamical resonance disappears if the coupling strengths to the $K\Lambda$ and $K\Sigma$ channels are reduced by about $40\%-50\%$.

A detailed study of the analytic structure of the reaction amplitudes has revealed the role of the $N^*(1535)$ and $N^*(1650)$ poles on other sheets; the pronounced cusp at the $\eta N$ threshold could be naturally explained. It appears as the result of an interplay of physical and hidden poles on different Riemann sheets. A virtual state in $S_{11}$ below the $\pi N$ threshold could be found that is quite stable and seems to be required by the sharp rise in Re $S_{11}$ that is seen in the partial wave analyses of, e.g., Ref. \cite{Arndt:2003if}. It remains to be seen if this state is genuine or mocked up from $t$- and $u$-channel cuts that are not explicitly included in the present model~\cite{privcom}. 

Furthermore, we have found that the $N^*(1535)$ pole disappears on some sheets; this implies a model-independent caveat: the use of sub-threshold resonances in phenomenological analyses is questionable. Thus, it is important to consider all relevant Riemann sheets and to pay special attention to which parts of the physical axis they are connected; sub-threshold and above-threshold poles, that are easily overlooked, have important consequences for the physical amplitude and should be considered.

While the present model describes the $S$ partial waves of the studied reactions well, there are residual discrepancies in the simultaneous description of the low and high energy regions. This could be traced back to the fact that the subtraction parameters are constants in energy; for a satisfactory description of the $S$-wave amplitudes covering the energy region from threshold to second resonance region, higher orders in the chiral meson-baryon interaction should be considered. Furthermore, although small for the $S_{11}$ amplitude, the $\pi\pi N$ channel should be included in the model.

We have shown that the existing $\pi N$ data do not rule out the description of the $N^*(1535)$ as a dynamically generated resonance. However, whether or not this scenario is indeed the case, still remains to be seen. In particular, the influence of the strong couplings to the $K\Lambda$ and $K\Sigma$ channels --which are responsible for the dynamical generation of this resonance-- on the higher partial waves should be investigated as mentioned in the Introduction.

While we have obtained a fair data description of the $\pi N$ final state in the present investigation with a dynamically generated $N^*(1535)$, the next logical step is to study the other final states ($\eta N$, $K\Lambda$, and $K\Sigma$) -- already included in the model as intermediate states -- in pion- and photon-induced reactions as well as in electroproduction to put further constraints on the model.

\vspace*{0.3cm}

\noindent {\bf Acknowledgements:} 
This work is supported by DFG (Deutsche Forschungsgemeinschaft, Gz: DO 1302/1-1) and the COSY FFE grant No. 41445282  (COSY-58). The authors are grateful to R.~A.~Arndt and I.~I. Strakovsky for discussions and J.~Haidenbauer and U.~G.~Mei{\ss}ner for a careful reading of the manuscript.

\end{document}